\documentclass[modern]{aastex631}

\usepackage{amsmath, amssymb, bm}
\usepackage{graphicx}
\usepackage[noabbrev]{cleveref}
\usepackage{appendix}
\usepackage{rotating}

\DeclareMathOperator*{\argmin}{arg\,min}

\DeclareMathOperator{\E}{\mathbb{E}}
\newcommand{\lr}[1]{\left( #1 \right)}
\newcommand{\tn}[1]{\textnormal{#1}}

\begin{document}

\title{Measuring the magnetic dipole moment and magnetospheric fluctuations
of accretion-powered pulsars in the Small Magellanic Cloud with an unscented Kalman filter}

\author[0000-0002-6547-2039]{Joseph O'Leary}
\affiliation{School of Physics, University of Melbourne, Parkville, VIC 3010, Australia.}
\affiliation{Australian Research Council Centre of Excellence for Gravitational Wave Discovery (OzGrav), Parkville, VIC 3010, Australia.}
\author{Andrew Melatos}
\affiliation{School of Physics, University of Melbourne, Parkville, VIC 3010, Australia.}
\affiliation{Australian Research Council Centre of Excellence for Gravitational Wave Discovery (OzGrav), Parkville, VIC 3010, Australia.}
\author{Tom Kimpson}
\affiliation{School of Physics, University of Melbourne, Parkville, VIC 3010, Australia.}
\affiliation{Australian Research Council Centre of Excellence for Gravitational Wave Discovery (OzGrav), Parkville, VIC 3010, Australia.}
\author{Nicholas J. O'Neill}
\affiliation{School of Physics, University of Melbourne, Parkville, VIC 3010, Australia.}
\affiliation{Australian Research Council Centre of Excellence for Gravitational Wave Discovery (OzGrav), Parkville, VIC 3010, Australia.}
\author{Patrick M. Meyers}
\affiliation{Theoretical Astrophysics Group, California Institute of Technology, Pasadena, CA 91125, USA.}
\author{Dimitris M. Christodoulou}
\affiliation{Lowell Centre for Space Science and Technology, Lowell, MA 01854, USA.}
\author{Sayantan Bhattacharya}
\affiliation{Lowell Centre for Space Science and Technology, Lowell, MA 01854, USA.}
\author{Silas G.T. Laycock}
\affiliation{Lowell Centre for Space Science and Technology, Lowell, MA 01854, USA.}
\affiliation{Department of Physics and Applied Physics, University of Massachusetts, Lowell, MA 01854, USA.}

\begin{abstract}
Many accretion-powered pulsars rotate in magnetocentrifugal disequilibrium, spinning up or down secularly over multi-year intervals. The magnetic dipole moment $\mu$ 
of such systems cannot be inferred uniquely from the time-averaged aperiodic X-ray flux $\langle L(t) \rangle$ and pulse period $\langle P(t) \rangle$, because the radiative efficiency of the accretion is unknown and degenerate with the mass accretion rate. Here we circumvent the degeneracy by tracking the fluctuations in the unaveraged time series $L(t)$ and $P(t)$ using an unscented Kalman filter, whereupon $\mu$ can be  estimated uniquely, up to the uncertainties in the mass, radius and distance of the star. The analysis is performed on \textit{Rossi X-ray Timing Explorer} observations for $24$ X-ray transients in the Small Magellanic Cloud, which have been monitored regularly for $\sim 16$ years. As well as independent estimates of $\mu$, the analysis yields time-resolved histories of the mass accretion rate and the Maxwell stress at the disk-magnetosphere boundary for each star, and hence auto- and cross-correlations involving the latter two state variables. The inferred fluctuation statistics convey important information about the complex accretion physics at the disk-magnetosphere boundary.
\end{abstract}


\section{Introduction}

The magnetic dipole moments $\mu$ of neutron stars span several orders of magnitude, with $10^{26} \lesssim \mu / (1 \, {\rm G \, cm^3}) \lesssim 10^{33}$ \citep{Lyne_2012}. For rotation-powered objects, $\mu$ is inferred from the spin-down rate measured by phase-connected pulse timing, assuming magnetic dipole braking \citep{Goldreich_1969,Ostriker_1969}. For accretion-powered objects, it is harder to measure $\mu$ accurately \citep{Mukherjee_2015}, because the relation between the spin-down rate and $\mu$ is more complicated for magnetocentrifugal accretion than for magnetic dipole braking. Resonant electron cyclotron lines yield direct estimates of the magnetic field strength in the line formation region near the stellar surface \citep{Makishima_1999,Makishima_2003,Caballero_2012,Revnivtsev_2016,Konar_2017,Staubert_2019} but they are detected in few objects. Accordingly, one usually resorts to inferring $\mu$ indirectly, e.g.\ by combining time-averaged X-ray timing data \citep{Ho_2014,Klus_2014,Mukherjee_2015} with physical theories of accretion; see Figure 6 in \cite{Klus_2014} and Figure 22 in \cite{Dangelo_2017} for example.

The Small Magellanic Cloud (SMC)  \citep{Corbet_2003} hosts more than $120$ high-mass X-ray binaries (HMXBs) \citep{Haberl_2022}; see \cite{White_1995} and \cite{Reig_2011} for overviews of X-ray binaries. The optical counterpart in each system is usually a Be star, with only one system hosting a supergiant donor star, namely SMC X$-$1 \citep{Haberl_2016}. The inferred distribution of $\mu$ for accretion-powered pulsars in the SMC covers a wide range, with $10^{29} \lesssim \mu / ( 1 \, {\rm G \, cm^3} ) \lesssim  10^{33}$ \citep{Klus_2014,Dangelo_2017}.  In two objects, $\mu$ is measured using cyclotron resonant scattering features, namely SXP 15.3 \citep{Maitra_2018} and SXP 2.37 (also known as SMC X$-$2) \citep{jaisawal_2016}; see \cite{Staubert_2019} for a detailed summary of X-ray sources showing cyclotron features in their spectra. In other objects, one combines temporal averages of the aperiodic X-ray flux $\langle L(t) \rangle$ and pulse period $\langle P(t) \rangle$ with assumptions about the accretion process, e.g.\ magnetocentrifugal equilibrium. However, one must be careful with the latter approach for the following three reasons. (i) The estimates depend on an approximate, phenomenological, magnetocentrifugal torque law; see Figure 7 in \cite{Shi_2015} and Figure 22 in \cite{Dangelo_2017} for a comparison of $\mu$ estimates for X-ray pulsars using different torque models. (ii) Some SMC X-ray pulsars are classified as rotating in a state of magnetocentrifugal disequilibrium; see Table 3 in \cite{Yang_2017} for an overview of the SMC X-ray pulsar population torque distribution. (iii) The radiative efficiency of the accretion is unknown; it is common to assume arbitrarily that $100\%$ of the gravitational potential energy of material falling onto the stellar surface is converted to heat and hence X-rays, but the assumption has not been verified independently. The time-averaged observables such as $\langle L(t) \rangle$ and $\langle P(t) \rangle$ do not contain enough independent pieces of information to infer $\mu$ uniquely, because the radiative efficiency is unknown. 

In this paper, we generalise previous magnetocentrifugal estimates of $\mu$ for accretion-powered pulsars in the SMC by exploiting the additional, time-dependent information in the fluctuations of $P(t)$ and $L(t)$, lifting the degeneracy that exists between $\mu$ and the radiative efficiency. We apply the Kalman filter parameter estimation framework developed in \cite{Melatos_2022} to (i) track the evolution of two hidden state variables associated with magnetocentrifugal accretion, namely the mass accretion rate and the Maxwell stress at the disk-magnetosphere boundary, which are important physically in their own right; and (ii) maximize the Kalman filter likelihood to infer the underlying, static, magnetocentrifugal parameters, including $\mu$. The analysis framework was validated by \cite{OLeary_2023} for the SMC X-ray transient SXP 18.3 using \textit{Rossi X-ray Timing Explorer} (RXTE) data \citep{Yang_2017}, yielding the first independent measurement of $\mu$ based on a Kalman filter. Here, we extend the analysis in \cite{OLeary_2023} to the entire catalogue of SMC X-ray pulsars in \cite{Yang_2017}. To do so, we replace the linear Kalman filter in \cite{Melatos_2022} with a nonlinear sigma-point Kalman filter \citep{Wan_2000,Wan_2001,Challa_2011}, which is suitable for all systems whether they are in equilibrium or disequilibrium. The above approach preserves more statistical information (e.g.,\ cross-correlations) than computing ensemble-averaged $P(t)$ and $L(t)$ statistics separately (e.g.,\ power spectral densities) \citep{Bildsten_1997,Riggio_2008,Klus_2014,Ho_2014,Mukherjee_2015,Serim_2023}. 

The paper is organized as follows. In Section \ref{Sec:MeasuringMagMom} we introduce the nonlinear, stochastic differential equations of motion which govern how the state variables associated with magnetocentrifugal accretion evolve \citep{Ghosh_1977,Ghosh_1979}, as well as the measurement equations which map the state variables, some of which are hidden, to the aperiodic X-ray flux $L(t)$ and the pulse period $P(t)$. The unscented Kalman filter and nested sampling algorithms used to infer the underlying, static, magnetocentrifugal parameters, including $\mu$, are introduced briefly in Section \ref{Sec:UKF}. The RXTE time series analyzed in this paper are discussed in Section \ref{Sec:HMXBSMC}. We extend the analysis to 24 SMC X-ray pulsars analyzed by \cite{Yang_2017} and present new estimates for (i) the underlying, static, physical parameters associated with magnetocentrifugal accretion in Section \ref{Sec:MagPE}; (ii) the magnetic dipole moment $\mu$ in Section \ref{Sec:MagMoms}; (iii) the radiative efficiency of the accretion in Section \ref{Sec:eta}; and (iv) the auto- and cross-correlation coefficients involving the mass accretion rate and the Maxwell stress at the disk-magnetosphere boundary in Section \ref{Sec:Corr}. Astrophysical implications are canvassed briefly in Section \ref{Sec:Conc} together with a note on generalizing the parameter estimation framework to persistent X-ray pulsars in the Milky Way. The unscented Kalman filter \citep{Wan_2000,Challa_2011} for nonlinear state and parameter estimation problems is summarised in Appendix \ref{AppA:UKF}. An end-to-end analysis of a single, representative pulsar, namely SXP $4.78$, is presented in Appendix \ref{App:WorkedExample} for the convenience of the reader as a worked example, to illustrate the steps involved and the output from each step.

\section{Measuring the magnetic moment}\label{Sec:MeasuringMagMom}

Accretion disks in HMXBs form due to mass transfer from a stellar companion via Roche lobe overflow or stellar winds. Disk accretion onto a magnetized, compact object is a time-dependent process, driven by complex hydromagnetic processes at the disk-magnetosphere boundary, including magnetorotational \citep{Balbus_1991}, Rayleigh-Taylor \citep{Stone_2007,Kulkarni_2008}, and Kelvin-Helmholtz \citep{Anzer_1980} instabilities. Three-dimensional magnetohydrodynamic simulations reveal complicated disk-magnetosphere interactions, mediated by twisted magnetic field lines and magnetic reconnection \citep{Romanova_2003,Romanova_2005,Kulkarni_2008,Fromang_2009,Romanova_2015,Das_2022}. Accordingly, the scalar observables $P(t)$ and $L(t)$ do not contain enough information to infer the spatial structure in the simulations, e.g.\ the geometry of the magnetic field at the disk-magnetosphere boundary. We therefore model accretion within the canonical, spatially averaged, magnetocentrifugal paradigm \citep{Ghosh_1979}, motivated by the promising results presented in \cite{Melatos_2022} and \cite{OLeary_2023}. In Section \ref{SubSec:ObsStateVariables}, we relate the observables $P(t)$ and $L(t)$ to the state variables associated with magnetocentrifugal accretion. The canonical magnetocentrifugal torque law, as well as a phenomenological, idealized model of the stochastic driving forces associated with hydromagnetic instabilities at the disk-magnetosphere boundary, are presented in Sections \ref{SubSec:MagnetoTorque} and \ref{SubSec:Stochastic}, respectively.  An explicit formula to estimate $\mu$ from the output of the Kalman filter and nested sampler conditional on the model in Sections \ref{SubSec:ObsStateVariables}--\ref{SubSec:Stochastic} is presented in Section \ref{SubSec:MagMomentObs}.  We refer the reader to \cite{Melatos_2022} for a detailed overview of the parameter estimation framework, and to \cite{OLeary_2023} for a practical guide on applying it to real astronomical data. 

\subsection{Observables and state variables}\label{SubSec:ObsStateVariables}

X-ray timing experiments with RXTE \citep{Levine_1996} collect raw photon counts, which are barycenter corrected and converted to $N$ samples of the pulse period $P(t_1),\hdots,P(t_N)$ and aperiodic X-ray luminosity $L(t_1),\hdots,L(t_N)$, using standard techniques and tools from X-ray timing analysis; see Section 2.3 in \cite{Yang_2017} for a practical guide.\footnote{The distance $D = 62 \pm 0.3 \, \rm{kpc}$ to the SMC is known to within $\pm 0.5\%$ \citep{Scowcroft_2016}, so we elect to work with $L(t)$ instead of the aperiodic X-ray flux.}  

Reformulated slightly from its original presentation, we express the standard magnetocentrifugal model of disk accretion \citep{Ghosh_1979} in terms of four time-dependent state variables: $\Omega(t)$, the angular velocity of the star; $Q(t)$, the rate at which matter flows from the accretion disk into the disk-magnetosphere boundary $\lr{\textnormal{units: } \textnormal{g} \; \textnormal{s}^{-1}}$; $S(t)$, the Maxwell stress at the disk-magnetosphere boundary $\lr{\textnormal{units: } \textnormal{g} \; \textnormal{cm}^{-1} \; \textnormal{s}^{-2}}$; and $\eta(t)$, the dimensionless radiative efficiency with which the gravitational potential energy of material falling onto the star is converted into X-rays. 

The state variables are not measured directly. They are related indirectly to the observables $P(t)$ and $L(t)$ through algebraic relations, which are nonlinear in general. Specifically one has
\begin{equation}\label{Eq:SpinPeriodMain}
    P(t) = 2 \pi / \Omega(t) + N_P(t),
\end{equation}
and
\begin{equation}\label{Eq:LuminosityMain}
    L(t) = GM Q(t) \eta(t) /R + N_L(t), 
\end{equation}
where Newton's gravitational constant and the mass and radius of the star are denoted by $G, M,$ and $R$ respectively. In this paper, we assume that  $N_P(t)$ and $N_L(t)$ are Gaussian noise processes, which satisfy the following ensemble statistics: $\langle N_P(t_n)\rangle = 0$, $\langle N_L(t_n) \rangle = 0$, $\langle N_P(t_n) N_P(t_{n'}) \rangle = \Sigma_{PP}^2 \, \delta_{n,n'}$, $\langle N_L(t_n), N_L(t_{n'})  \rangle = \Sigma_{LL}^2 \, \delta_{n,n'}$, and $\langle N_P(t_n), N_L(t_{n'})\rangle = 0$, where $\delta_{n,n'}$ denotes the Kronecker delta. The latter assumption may not hold, if photon counting and time tagging are correlated operations in the RXTE detector. It can be relaxed, if future data warrant. 

It is possible in principle to reconstruct the time series for the four state variables $\Omega(t)$, $Q(t)$, $S(t)$, and $\eta(t)$ from the two observables $P(t)$ and $L(t)$ conditional on the dynamical model defined in Sections \ref{SubSec:MagnetoTorque} and \ref{SubSec:Stochastic}. This property, known as identifiability in the electrical engineering literature \citep{bellman_1970}, is proved formally in Appendix D in \cite{Melatos_2022}. The state variable $\Omega(t)$ in Equation (\ref{Eq:SpinPeriodMain}) is related to $Q(t)$ and $S(t)$ via the canonical magnetocentrifugal torque law in Section \ref{SubSec:MagnetoTorque}. A phenomenological, idealized model for the $Q(t)$ and $S(t)$ dynamics is stipulated in Section \ref{SubSec:Stochastic}.  The radiative efficiency $0 < \eta(t) < 1$ in Equation (\ref{Eq:LuminosityMain}) is the product of two factors: (i) the fraction of specific gravitational potential energy $GM/R$ of material falling onto the stellar surface that is converted to heat and hence X-rays, e.g.\ via bremsstrahlung emission; and (ii) the fraction of $Q(t)$ which strikes the stellar surface, and does not escape through outflows \citep{Matt_2005,Matt_2008,Romanova_2015,Marino_2019}. As discussed in Appendix C in \cite{Melatos_2022} and Section 2.1 in \cite{OLeary_2023}, more realistic models require more than the $\sim 10^3$ samples available per object analyzed in this paper to disentangle factors (i) an (ii) above \citep{Dangelo_2010,Dangelo_2012,Dangelo_2017}.  Accordingly, we stick with the single scalar variable $\eta(t)$ and make the additional simplifying assumption $\eta(t) = \bar{\eta} = {\rm constant}$, i.e.\ we hold $\eta(t)$ constant in time but leave its value free to be estimated by the Kalman filter. Even with the latter assumption, one can estimate $\bar{\eta}$ and hence $\mu$ independently, the chief goal of this paper. 

\subsection{Magnetocentrifugal torque}\label{SubSec:MagnetoTorque}

In the standard model of magnetocentrifugal disk accretion \citep{Ghosh_1977}, the accretion torque is governed by the location of the disk-magnetosphere boundary, which is assumed to have zero thickness, and occurs at the Alfv\'en radius $R_{\rm m}(t)$. It is located where the disk ram pressure balances the Maxwell stress $S(t)$, viz.  $S \approx \rho v^2/2$, where $\rho = Q/\lr{4\pi R_{\rm m}^2 v}$ and $v = \lr{2 GM/R_{\rm m}}^{1/2}$ are the mass density and infall speed respectively in free fall \citep{Menou_1999,Frank_2002}.\footnote{The effective infall speed $v$ differs by a factor of order unity in different models and geometries, e.g.\ \cite{Melatos_2022} take $v \approx (GM/R_{\rm m})^{1/2}$. \label{FN:infallSpeed}} Thus, the Alfv\'en radius is given in terms of the state variables by 
\begin{equation}\label{Eq:AlfvenRad}
    R_{\rm m} (t) = \lr{2 \pi^{2/5}}^{-1} \lr{GM}^{1/5} Q(t)^{2/5} S(t)^{-2/5}.
\end{equation}
The corotation radius, 
\begin{equation}\label{Eq:CoRotRad}
R_{\rm c}(t) = \lr{GM}^{1/3}\Omega\lr{t}^{-2/3},
\end{equation}
is located where the stellar rotation rate equals the Kepler frequency of the disk material. The fastness parameter $(R_{\rm m}/R_{\rm c})^{3/2}$ governs the sign of the magnetocentrifugal torque. For $R_{\rm m} < R_{\rm c}$, material strikes the stellar surface, causing the star to spin up, with $\tn{d}P/\tn{d}t < 0$. For $R_{\rm m} > R_{\rm c}$, the material at the disk-magnetosphere boundary orbits slower than the star, with matter being ejected centrifugally by the corotating magnetosphere, causing the star to spin down, with $\tn{d}P/\tn{d}t > 0$. For $R_{\rm m} \approx R_{\rm c}$, the star is in rotational equilibrium, i.e.\ there is zero net torque, with  $\tn{d}P/\tn{d}t \approx 0$. In this paper we consider systems in equilibrium and disequilibrium, as classified in Table 3 in \cite{Yang_2017}. The net torque acting on the star involves both hydromagnetic and dynamical contributions and is given by 
\begin{equation} \label{Eq:SpinEquation}
    I \frac{\tn{d} \Omega}{\tn{d}t} = \lr{GM}^{1/2} \left\{ 1 -  \left[\frac{R_{\rm m}\lr{t}}{R_{\rm c}\lr{t}}\right]^{3/2} \right\} R_{\rm m}\lr{t}^{1/2}Q\lr{t},
\end{equation}
where the star's moment of inertia is denoted by $I$. Here we neglect modifications to Equation (\ref{Eq:SpinEquation}), e.g.\ due to gravitational radiation reaction \citep{bildsten_1998,Melatos_2005}, as the data are not detailed enough to warrant their inclusion. 

The torque law (\ref{Eq:SpinEquation}) does not capture fully several important time-dependent phenomena observed in accretion-powered pulsars, e.g.\ the rapid onset of the propeller transition, episodic accretion, and trapped disk scenarios \citep{Dangelo_2010,Dangelo_2012,Dangelo_2015,Dangelo_2017}. Current data volumes with \ $\sim 10^3$ samples are too small to expose corrections to Equation (\ref{Eq:SpinEquation}), but a Kalman filter framework can be developed for studying such corrections with more data in the future \citep{Dangelo_2015,Dangelo_2017,Melatos_2022}. Specifically, we refer the reader to Appendix C.1 in \cite{Melatos_2022} and Section 2.2 in \cite{OLeary_2023} for discussions about refining the complexities at the disk-magnetosphere boundary and to Appendices C.2 and C.3 in \cite{Melatos_2022} for a guide on implementing propeller transition and trapped disk models for parameter estimation problems with a nonlinear Kalman filter. 

\subsection{Stochastic fluctuations} \label{SubSec:Stochastic}

The stochastic variability in the measured pulse period $P(t)$ and aperiodic X-ray luminosity $L(t)$ for accretion-powered pulsars \citep{Bildsten_1997} is driven by several mechanisms, e.g.\ Kelvin-Helmholtz and Rayleigh-Taylor instabilities at the disk-magnetosphere boundary \citep{Romanova_2003,Romanova_2008,Romanova_2012}, and modulation in the accretion rate $Q(t)$ as the companion star evolves; see Section 2.4 in \cite{Melatos_2022} for more examples. 

In this paper, we assume that the state variables $Q(t)$ and $S(t)$ execute mean-reverting random walks driven by white noise \citep{deKool_1993}. That is, $Q(t)$ and $S(t)$ satisfy the Langevin equations
\begin{eqnarray}
    \frac{d Q}{d t} &= -\gamma_Q [Q(t) - \bar{Q}] + \xi_Q(t), \label{eq:QLangevin} \\
    \frac{d S}{d t} &= -\gamma_S [S(t) - \bar{S}] + \xi_S(t), \label{eq:SLangevin}
\end{eqnarray}
where $\gamma_Q$ and $\gamma_S$ denote damping constants, and $\xi_Q(t)$ and $\xi_S(t)$ are white-noise driving terms satisfying $\langle \xi_{Q}(t)\rangle = 0$, $\langle \xi_S(t) \rangle = 0$, $\langle \xi_Q(t) \, \xi_Q(t') \rangle = \sigma_{QQ}^2 \, \delta (t - t')$, $\langle \xi_S(t) \, \xi_S(t')  \rangle = \sigma_{SS}^2 \, \delta(t - t')$, and $\langle \xi_Q(t) \, \xi_S (t') \rangle = 0$.  Equations (\ref{eq:QLangevin}) and (\ref{eq:SLangevin}) ensure that $Q(t)$ and $S(t)$ wander randomly about their asymptotic ensemble-averaged values, denoted by $\bar{Q}$ and $\bar{S}$, with characteristic time scales of mean reversion $\gamma_Q^{-1}$ and $\gamma_S^{-1}$, and rms fluctuations $\sim \gamma_Q^{-1/2}\sigma_{QQ}$ and $\sim \gamma_S^{-1/2}\sigma_{SS}$, respectively. Note that $\bar{Q}$ and $\bar{S}$ do not necessarily correspond to magnetocentrifugal equilibrium; they can lead to $d \langle \Omega \rangle /dt \neq 0$ for example. Finally, the magnetocentrifugal torque law (\ref{Eq:SpinEquation}) and Langevin equations (\ref{eq:QLangevin}) and (\ref{eq:SLangevin}) are supplemented with the following deterministic equation for the radiative efficiency, justified in Section \ref{SubSec:ObsStateVariables}:
\begin{equation} \label{eq:EtaConstant}
    \eta(t) = \bar{\eta}.
\end{equation}

Many aspects of the complex fluctuation dynamics of accretion-powered pulsars, as observed in three-dimensional simulations \citep{Romanova_2003,Romanova_2005,Kulkarni_2008}, are not captured fully by Equations (\ref{eq:QLangevin})--(\ref{eq:EtaConstant}). In particular, the assumption $\eta(t)=\bar{\eta}=\rm{constant}$ is unlikely to hold exactly in reality, for reasons which are discussed in detail in Section 2.4 and Appendix C of \cite{Melatos_2022}, as well as Section 4.4 of \cite{OLeary_2023}. There is no consensus in the literature on how best to model the temporal evolution of the radiative efficiency in general and hence it is impossible to say at present, theoretically or observationally, whether real accretion-powered pulsars are approximated better by  Equation (\ref{eq:EtaConstant}), or (for example) the generalized Equations (B3) or (C2) in \cite{Melatos_2022}, because existing data are insufficient to select confidently between the many possible models.

One situation where $\eta(t) = \bar{\eta} = \rm{constant}$ may not hold exactly is the rapid onset of the propeller transition, in which the sign of $\tn{d}\Omega/\tn{d}t$ switches, when $R_{\rm m} - R_{\rm c}$ increases from $-\Delta R_2$ to $+\Delta R_2$ with $\Delta R_2 \ll R_{\rm c}$ \citep{Dangelo_2010}, unlike in Equation (\ref{Eq:SpinEquation}), where the sign of $\tn{d}\Omega/\tn{d}t$ switches, when $R_{\rm m}- R_{\rm c}$ increases from $-R_{\rm c}$ to $+R_{\rm c}$, where $\Delta R_2$ denotes the length scale with which material transitions between free fall and propeller; see Appendix C.2 of \cite{Melatos_2022}. The propeller onset is accompanied by a rapid drop in $\eta(t)$ to $\eta(t)\ll 1$, the qualitative features of which are observed in Figure 5 in \cite{Melatos_2022}. A second example is disk trapping in the weak propeller regime $R_{\rm c} \leq R_{\rm m} \lesssim 1.3 R_{\rm c}$, where the centrifugal force generated by the rotating magnetosphere is insufficient to gravitationally unbind matter from the disk. The complicated disk-magnetosphere interaction may deposit enough angular momentum to stall the inflow, so that gas piles up near $R_{\rm m}$ before it breaks through and accretes episodically onto the star, accompanied by spikes in $\eta(t)$ \citep{Dangelo_2010,Dangelo_2012,Dangelo_2017}, as observed in Figure 6 in \cite{Melatos_2022}. It is possible in principle to incorporate these effects into the Kalman filter framework in an idealized way, using phenomenological analytic formulas based on the hyperbolic tangent approximation of \cite{Dangelo_2017} for example. The reader is directed to Appendices C.2 and C.3 of \cite{Melatos_2022} for implementation details and validation tests with synthetic data.

Other dynamical variables in the model in Section \ref{Sec:MeasuringMagMom} are simplified, not just $\eta(t)$. (i) In the interaction region, $R_{\rm m} - \Delta R \leq r \leq R_{\rm m}$, the magnetic field is sheared due to differential rotation between the disk and star, and the torque on the star is proportional to the product of the toroidal and vertical components of the magnetic field, which evolve independently; see Equation (1) in \cite{Dangelo_2010}. The scalar Maxwell stress $S(t)$ does not capture fully the independent evolution of the toroidal and vertical field components \citep{Melatos_2022}. (ii) In regions where magnetic pressure dominates over gas pressure, the magnetic field lines tend to open, disconnecting the disk at $r \gtrsim R_{\rm m}$ magnetically from the star \citep{Dangelo_2010}. This modifies Equation (\ref{eq:SLangevin}) for $S(t)$.  (iii) If enough angular momentum and energy are deposited, parts of the disk at $r \gtrsim R_{\rm m}$ unbind, launching a vertical outflow along the twisted magnetic field lines \citep{Matt_2005, Romanova_2015, Melatos_2022}. Outflows modify Equations (\ref{eq:QLangevin}) and (\ref{eq:EtaConstant}) for $Q(t)$ and $\eta(t)$, respectively. Points (i)--(iii) compound the idealizations inherent in the torque law (\ref{Eq:SpinEquation}), discussed in Section 2.2, and the idealizations inherent in (\ref{eq:EtaConstant}) through the definition and constancy of $\eta(t)$, discussed in Section \ref{SubSec:ObsStateVariables}. Altogether, Equations (\ref{Eq:SpinEquation})--(\ref{eq:EtaConstant}) represent a compromise between realism and the explanatory power of the available volume of data. The data are sufficient to estimate six fundamental model parameters, as discussed in Section \ref{SubSec:MagMomentObs}, but are insufficient to estimate the additional parameters involved in more complicated physical models of $\eta(t)$, which would be needed to treat points (i)--(iii) more realistically. Examples of more complicated models can be found in Appendix C of \cite{Melatos_2022}.

\subsection{Magnetic dipole moment from the nested sampler }\label{SubSec:MagMomentObs} 

The parameters $(\bar{\Omega},\bar{Q},\bar{S},\bar{\eta})$ form a key input into the unscented Kalman filter in Section \ref{Sec:UKF}. They are related to the six magnetocentrifugal parameters estimated by the Kalman filter and nested sampler from the data, namely $\bm{\Theta} = (\beta_1, \beta_2, \gamma_Q, \gamma_S, \sigma_{QQ}, \sigma_{SS})$, with\footnote{Equations (\ref{eq:beta1}) and (\ref{eq:beta2}) differ from Equations (B4) and (B5) in \cite{Melatos_2022} by factors of order unity due to different interpretations of the infall speed $v$ in different models; see footnote \ref{FN:infallSpeed}.}
\begin{equation} \label{eq:beta1}
    \beta_1 = \frac{(GM)^{3/5} \bar{Q}^{6/5}}{(2\pi^{2/5})^{1/2}I \, \bar{\Omega} \, \bar{S}^{1/5}}
\end{equation}
and 
\begin{equation} \label{eq:beta2}
    \beta_2 = \frac{(GM)^{2/5}\bar{Q}^{9/5}}{(2 \pi^{2/5})^{2} \, I \, \bar{S}^{4/5}}.
\end{equation}
Equations (\ref{eq:beta1}) and (\ref{eq:beta2}) are deduced  by introducing the dimensionless time-dependent functions, $A_1(t) = A(t)/\bar{A}$ with $A \in \{\Omega, Q, S\}$, in Equations (\ref{Eq:AlfvenRad})--(\ref{Eq:SpinEquation}). An equivalent scaling, $A_1(t) = A(t)/\bar{A}$ with $A \in \{ P, L, \eta \}$, is applied to Equations (\ref{Eq:SpinPeriodMain}), (\ref{Eq:LuminosityMain}), and (\ref{eq:EtaConstant}). 

Upon introducing the scaling $A_1(t) = A(t)/\bar{A}$ for $A \in \{ \Omega, Q, S\}$ in Equations (\ref{Eq:SpinEquation})--(\ref{eq:SLangevin}),   six independent parameters remain, i.e.\ $\bm{\Theta} = (\beta_1, \beta_2, \gamma_Q, \gamma_S, \sigma_{QQ}, \sigma_{SS})$, termed the fundamental magnetocentrifugal parameters \citep{Melatos_2022,OLeary_2023}. The mean-reversion time scales, $\gamma_Q^{-1}$ and $\gamma_S^{-1}$, as well as the rms noise amplitudes, $\sigma_{QQ}$ and $\sigma_{SS}$, have clear physical interpretations stemming from Markov process theory \citep{Gardiner_1985}. The parameters, $\beta_1$ and $\beta_2$, are less obvious to interpret at first glance. Mathematically, $\beta_1$ and $\beta_2$ correspond to the coefficients of the spin-up and spin-down accretion torques defined on the right-hand side of Equation (\ref{Eq:SpinEquation}), respectively. Astrophysically, $\beta_1$ and $\beta_2$ are connected with the fastness parameter, viz.\ $\beta_2/\beta_1 = (\bar{R}_{\rm{m}}/\bar{R}_{\rm c})^{3/2}$ for $\bar{R}_{\rm{m}} = \lr{2 \pi^{2/5}}^{-1} \lr{GM}^{1/5} \bar{Q}^{2/5} \bar{S}^{-2/5}$ and  $\bar{R}_{\rm c} = \lr{GM}^{1/3}\bar{\Omega}^{-2/3}$. The implications of $\beta_1$ and $\beta_2$, e.g.\ those deduced from Figure \ref{fig:BetaDistribution}, are discussed briefly in Section \ref{Sec:MagPE} and will be explored further in a forthcoming paper.

Once $\bm{\Theta} = (\beta_1, \beta_2, \gamma_Q, \gamma_S, \sigma_{QQ}, \sigma_{SS})$ is estimated by the Kalman filter and nested sampler, we solve Equations (\ref{eq:beta1}) and (\ref{eq:beta2}) simultaneously for $\bar{Q}$ and $\bar{S}$, while $\bar{\Omega} = \langle \Omega(t) \rangle$ is determined directly from the data by calculating the sample average 
\begin{equation} \label{Eq:SampleMeansOmega}
    \bar{\Omega} = N^{-1} \sum^{N}_{n=1} 2 \pi / P(t_n).
\end{equation}
The average of the radiative efficiency $\bar{\eta}$ is then calculated by solving\footnote{In  a more general accretion model, where fluctuations in $\eta(t)$ are modelled using (for example) Equation (B3) in \cite{Melatos_2022}, $Q(t)$ and $\eta(t)$ are correlated, and $\beta_3 = \langle Q(t) \, \eta(t) \rangle/[\langle Q(t) \rangle \, \langle \eta(t)\rangle]$ represents an additional parameter to be estimated by the Kalman filter; see Appendix B2 in \cite{Melatos_2022}.}
\begin{equation} \label{eq:AveragedLstate}
    \bar{L} = GM \bar{Q}\bar{\eta}/R,
\end{equation}
for $\bar{\eta}$, where $\bar{L}$ is computed from the sample mean
\begin{equation}\label{Eq:SampleMeansL}
    \bar{L} = N^{-1}\sum^{N}_{n=1}L(t_n).
\end{equation}
The magnetic dipole moment $\mu$ is then expressed in terms of $\bar{Q}$ and $\bar{S}$, via $\mu = (2\pi \, \bar{S})^{1/2} \, \bar{R}_{\rm m}^{3}$, as follows:
\begin{equation} \label{eq:MuEquation}
    \mu = 2^{-5/2}\pi^{-7/10} (GM)^{3/5} \bar{Q}^{6/5}\bar{S}^{-7/10}.
\end{equation} 

The barred quantities $\bar{Q}$ and $\bar{S}$, which satisfy $\bar{Q} \approx \langle Q(t) \rangle$ and $\bar{S} \approx \langle S(t) \rangle$ asymptotically for $t \gg \max \{\gamma_Q^{-1},\gamma_S^{-1} \}$, appear in Equation (\ref{eq:MuEquation}) because of the following physical subtlety. Strictly speaking, the system satisfies $\mu = (2\pi \bar{S})^{1/2} \bar{R}_{\rm m}^3$ but not $\mu = [2\pi S(t)]^{1/2} R_{\rm m}(t)^3$, because Equation (\ref{eq:SLangevin}) drives white-noise fluctuations in $S(t)$ independently from fluctuations in $R_{\rm m}(t)$ in general. For example, instability-driven fluctuations in the magnetic geometry at the disk-magnetosphere boundary occur in principle even when $R_{\rm m}$ is constant, although in practice $R_{\rm m}(t)$ fluctuates too. For the same reason, one cannot write $\mu \propto Q(t)^{6/5} S(t)^{-7/10}$ or $\mu \propto \langle Q(t)^{6/5} S(t)^{-7/10} \rangle$ to replace Equation (\ref{eq:MuEquation}) in general, whereas replacements of this kind are possible in the original, noiseless, magnetocentrifugal picture \citep{Ghosh_1977}.

\section{Parameter estimation via a Kalman filter}\label{Sec:UKF}

Parameter estimation occurs in two steps in this paper. In step one, an unscented Kalman filter \citep{Wan_2000,Challa_2011} is applied to the nonlinear measurement equations (\ref{Eq:SpinPeriodMain}) and (\ref{Eq:LuminosityMain}) and equations of motion (\ref{Eq:SpinEquation})--(\ref{eq:EtaConstant}) to infer the optimal state trajectory $\bm{X}_n = [\Omega(t_n),Q(t_n),S(t_n),\eta(t_n)]$ given the measurements $\bm{Y}_n=[P(t_n),L(t_n)]$ for $1\leq n \leq N$ at fixed, arbitrary $\bm{\Theta}$, where $N$ denotes the number of samples per pulsar. The number $N$ of $P(t_n)$ and $L(t_n)$ samples for each object is listed in the eighth column of Table 2 in \cite{Yang_2017}; see also Figure \ref{fig:ObsDistribution} for a summary of the number of samples associated with the HMXBs analyzed in this paper.  In step two, the static parameters $\bm{\Theta} = (\beta_1,\beta_2,\gamma_Q,\gamma_S,\sigma_{QQ},\sigma_{SS})$ are allowed to vary, and a nested sampler \citep{Skilling_2004,Ashton_2022} is employed to find the value of $\bm{\Theta}$ that maximizes the Bayesian likelihood of the data given the model. The framework is suitable for systems which rotate in equilibrium or disequilibrium.  The Kalman filter likelihood and state update recursion relations are summarized briefly in this section, together with an overview of the nested sampling algorithm.  We refer the reader to Appendix \ref{AppA:UKF} for a fuller summary of the unscented Kalman filter and nested sampler used in this paper. 

The log-likelihood of a Kalman filter takes the form \citep{Meyers_2021}
\begin{equation} \label{eq:loglikelihoodmain}
    \ln{p\lr{ \{\bm{Y}_n\}_{n=1}^N |\bm{\Theta}}} = -\frac{1}{2}\sum^N_{n=1}\left[ D_{\bm{Y}} \ln{\lr{2\pi}} + \ln{\textnormal{det}\lr{\bm{s}_n}} + \bm{e}_n^T \bm{s}_n^{-1} \bm{e}_n \right],
\end{equation}  
where $D_{\bm{Y}} = 2$ denotes the dimension of $\bm{Y}_n$, $\bm{e}_n$ is the innovation vector, and $\bm{s}_n$ is the predicted measurement (innovation) covariance, defined according to Equation $(\ref{Eq:InnovationCovariance})$.  At every time step $t_n$ the unscented Kalman filter calculates the pre-fit innovation vector according to 
\begin{equation} \label{eq:innovation}
    \bm{e}_n = \bm{Y}_n - \bm{Y}_n^{-},
\end{equation}
where $\bm{Y}^{-}_{n}$ denotes the predicted measurement. The state vector is updated recursively via 
\begin{equation} \label{eq:stateUpdate}
    \bm{X}_n = \bm{X}_n^{-} + \bm{k}_n \bm{e}_n,
\end{equation}
where $\bm{X}_n^{-}$ and $\bm{k}_n$ denote the predicted state and Kalman gain respectively. Equations (\ref{eq:innovation}) and (\ref{eq:stateUpdate}) are standard formulas in the context of nonlinear unscented Kalman filters and are reproduced here for overall context. We refer the reader to Appendix \ref{AppA:UKF} for an overview regarding their structure and implementation. 

In this paper, we evaluate Equation (\ref{eq:loglikelihoodmain}) as a function of $\bm{\Theta}$ using the \texttt{dynesty} nested sampler \citep{Speagle_2020} through the \texttt{bilby} \citep{Ashton_2019} front-end. In its simplest form, nested sampling is a computational tool in the context of Bayesian inference for computing marginal likelihoods and parameter posterior distributions. It proceeds as follows. The sampler is initialized with $N_{\rm live}$ ``live'' points drawn randomly from the prior $p(\bm{\Theta})$, denoted by $\bm{\Theta}^{(1)}_{1}, \hdots,\bm{\Theta}^{(1)}_{N_{\rm live}}$. At every step $k$, and for all $1 \leq m \leq N_{\rm live}$, the sampler calculates $\mathcal{L}_{m}^{(k)} = \ln{p [\{\bm{Y}_n\}_{n=1}^N|\bm{\Theta}^{(k)}_m]}$, and replaces the live point whose likelihood is lowest, namely $\bm{\Theta}^{(k)}_{m'}$ with $m' = \argmin_{m} \mathcal{L}_m^{(k)}$, with a new live point $\bm{\Theta}^{(k+1)}_{m'}$ drawn from the prior, subject to the condition $\mathcal{L}_{m'}^{(k+1)} > \mathcal{L}_{m'}^{(k)}$. The process repeats until a suitable stopping condition is met. The above outline follows closely the discussion in Section 4.1 in \cite{OLeary_2023}. We refer the reader to Sections 5 and 6 in \cite{Skilling_2006} for a detailed overview of the nested sampling algorithm. 

\section{RXTE observations of SMC X-ray pulsars}\label{Sec:HMXBSMC}

The Proportional Counter Array (PCA) \citep{Jahoda_2006} is the primary X-ray timing instrument onboard the RXTE spacecraft. For $\sim$ 16 years, it monitored the SMC regularly, collecting timing data in the 2$-$60 keV band with submillisecond timing resolution using four Xenon-filled Proportional Counter Units (PCUs) \citep{Revnivtsev_2003, Yang_2017}. RXTE timing data is archived and publicly available via the High Energy Astrophysics Science Archive Research Centre.\footnote{\url{https://heasarc.gsfc.nasa.gov/}} 
Post-processed RXTE data products have been instrumental in estimating the orbital and stellar properties of accretion-powered pulsars in the SMC \citep{Galache_2008,Townsend_2011,Klus_2014,Coe_2015,Christodoulou_2017}, as well as constraining black hole spin parameters among other applications \citep{Shafee_2005,Remillard_2006,Reynolds_2021}. 
In Section \ref{SubSec:RXTEData} we summarise the acquisition of the $P(t_n)$ and $L(t_n)$ time series used in this paper.
The spin distribution of HMXBs in the SMC, as classified in Table 3 in \cite{Yang_2017}, is discussed in Section \ref{SubSec:RotState}.  
\subsection{$P(t_n)$ and $L(t_n)$ time series}\label{SubSec:RXTEData}
The RXTE PCA records photon time-of-arrivals, which are subsequently converted to light curves using standard X-ray timing techniques, i.e.\ they are barycenter corrected with background noise subtracted; see Sections 1.3 and 4.4.1 in \cite{Laycock_2005} and \cite{Patruno_2021} respectively for overviews on X-ray timing data analysis.

\begin{figure}
\centering
\includegraphics[width=1\textwidth, keepaspectratio]{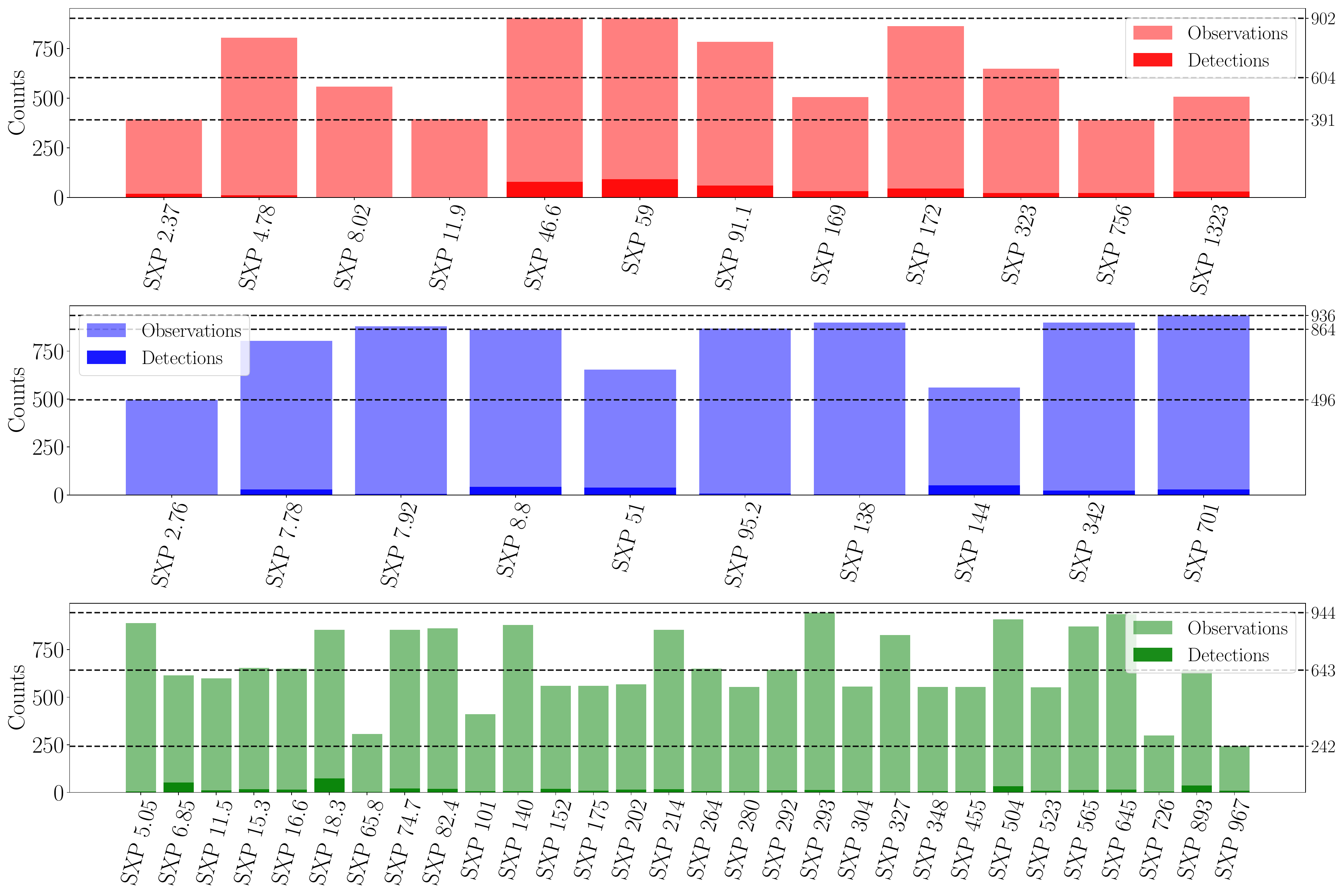}
        \caption{RXTE PCA aperiodic X-ray flux measurements collected between 1995 and 2012 for 52 SMC HMXBs \citep{Yang_2017}. The first (red bars), second (blue bars), and third (green bars) panels correspond to stars classified as spinning up, down, and in rotational equilibrium, respectively by \cite{Yang_2017}. The label ``Observations'' refers to the total number of measurements collected over the observing period and includes measurements that correspond to episodes of quiescence and periods of outburst. The label ``Detections'' refers to statistically significant X-ray timing points (at the $99\%$ level), as classified by \cite{Yang_2017}. The minimum, median, and maximum values associated with the total number of observations in each panel are overplotted as dashed, horizontal lines. }
    \label{fig:ObsDistribution}
\end{figure}

In this paper, we employ the post-processed pulse period and aperiodic X-ray luminosity time series presented in \cite{Yang_2017}. The latter authors analyze a total of 36316 RXTE X-ray timing observations of the SMC, and generate light curves in the 3--10 keV energy range. They search the light curves at the fundamental and harmonic frequencies of the known SMC HMXBs in the PCA field of view and estimate the spin period and pulse amplitude for each X-ray timing observation by Lomb-Scargle methods; see Section 2.4 in \cite{Yang_2017} for further details about RXTE data processing techniques, as well as \cite{VanderPlas_2018} for a practical guide to extracting periodic signals from irregularly sampled time series with Lomb-Scargle periodograms. The sample size $N$ per pulsar is determined by the number of times each object, whose sky position is  known a priori,  enters the RXTE PCA field of view with collimator response $> 0.2$ \citep{Yang_2017}. The 36316 X-ray timing observations vary in duration, and are tagged with a unique observation identification number \texttt{ObsID}, with each \texttt{ObsID} corresponding to one post-processed $P(t_n)$ sample. The library published by \cite{Yang_2017} contains 52 pulse period time series whose dimension, i.e.\ the number of $P(t_n)$ samples per object, is reported in Table 2 in \cite{Yang_2017} and summarized in Figure \ref{fig:ObsDistribution} below for the convenience of the reader.

The PCA is a non-imaging detector. Several X-ray sources regularly occupy its field of view. In order to avoid source confusion, the aperiodic X-ray flux is not measured directly. Rather, it is estimated empirically from the measured pulse amplitude with the Portable, Interactive Multi-Mission Simulator (PIMMS) toolkit.\footnote{\url{https://cxc.harvard.edu/toolkit/pimms.jsp}} Using PIMMS power-law modelling tools, \cite{Yang_2017} estimated the total flux for 1 count $\rm{PCU^{-1} s^{-1}}$ as $9.23^{+0.10}_{-0.12} \times 10^{-12} \, \rm{erg \, cm^{-2} \, s^{-1}}$ and approximated the aperiodic X-ray luminosity assuming a fixed pulsed fraction of $0.4$; see Section 2.4.2 and Equation (5) in \cite{Yang_2017} for details on estimating luminosity using pulse amplitude, pulsed fraction, and flux. The RXTE PCA flux measurements collected between 1995 and 2012 are summarized in Figure \ref{fig:ObsDistribution} for the convenience of the reader. The first, second, and third panels contain SMC X-ray transients classified as spinning up (red bars), down (blue bars), and in rotational equilibrium (green bars) respectively; see Table 3 in \cite{Yang_2017}.

We quantify the $P(t_n)$ and $L(t_n)$ uncertainty using the measurement noise covariance $\bm{\Sigma}_n$ at time $t_n$, discussed in detail in Appendix \ref{AppA:StateTracking}. \cite{Yang_2017} calculated the pulse period uncertainty per $P(t_n)$ sample using Equation (14) in \cite{Horne_1986}. Given that $L(t_n)$ is not measured directly by the PCA but is calculated empirically, we approximate the luminosity uncertainty per $L(t_n)$ sample by the variance of the $L(t_n)$ time series for each of the 52 objects analyzed in this paper. That is, for every object, every $P(t_n)$ sample has its own unique error bar calculated according to \cite{Horne_1986}, but every $L(t_n)$ sample for that object has the same error bar, calculated from the variance of $L(t_1), \hdots, L(t_N)$. We refer the reader to Section 2 in \cite{Horne_1986} for fuller details on estimating pulse period uncertainty and to Appendix \ref{AppA:StateTracking} for an overview on how the Kalman filter handles noisy measurements through the Kalman gain $\bm{k}_n$. We quantify the relation between the measurement uncertainties in $L(t_n)$ and $P(t_n)$ and the accuracy of the Kalman filter output in Section 4.2 of \cite{OLeary_2023}. Another study of the relation between the measurement uncertainties and the accuracy of the Kalman filter output is given in Appendix D.2 of \cite{ONeill_2024} for a related but different parameter estimation problem, namely analyzing radio pulsar timing noise with a Kalman filter.

\subsection{Rotational state: equilibrium or disequilibrium} \label{SubSec:RotState}

Many accretion-powered pulsars in the SMC and elsewhere exist well away from a state of rotational equilibrium, spinning up or down secularly over multi-year intervals, punctuated by sharp torque reversals \citep{Bildsten_1997, Yang_2017}; see Figure 1 in \cite{Serim_2023} for an example of a torque reversal in the low mass X-ray binary 4U 1626$-$67. While the linear Kalman filter analysis in \cite{Melatos_2022} and \cite{OLeary_2023} is suitable for systems classified as being near rotational equilibrium, i.e.\  $R_{\rm m}(t) = R_{\rm c}(t) = R_{\rm m 0} = \rm{constant}$, it is not suitable for systems in disequilibrium. In this paper, therefore, we elect to apply a nonlinear unscented Kalman filter to $52$ SMC X-ray sources, whose spin properties were determined in \cite{Yang_2017}, as described below.  

\begin{figure}
\centering{
    \includegraphics[width=\textwidth, keepaspectratio]{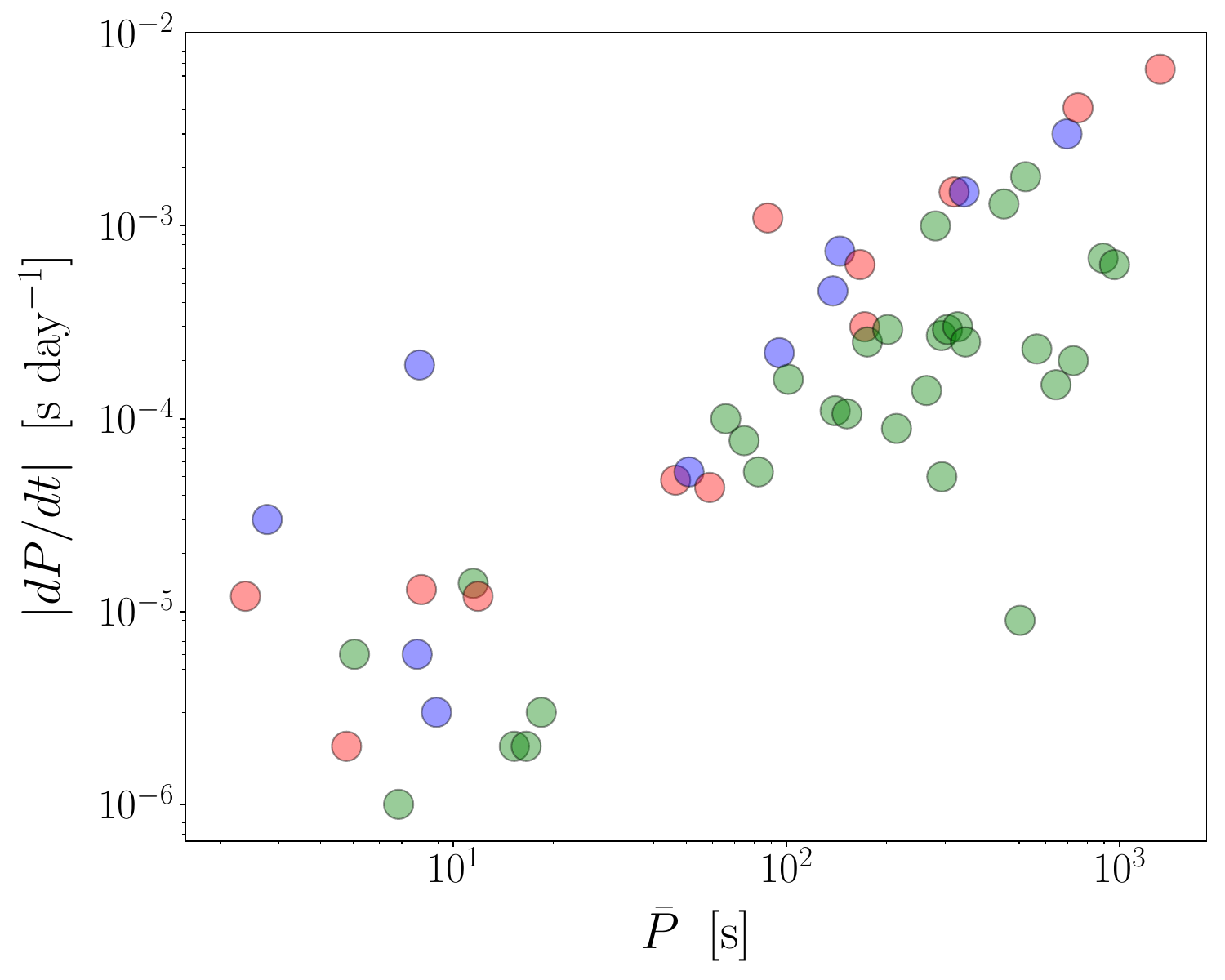}}
        \caption{$P$--$|\dot{P}|$ diagram of the X-ray pulsars analyzed by \cite{Yang_2017}. The red and blue points correspond to stars classified as spinning up and down respectively, while green points indicate rotational equilibrium, as classified in Table 3 in \cite{Yang_2017}. The horizontal and vertical axes are plotted on $\log_{10}$ scales.}
    \label{fig:SpinDistribution}
\end{figure}

In addition to publishing time series of the pulse period, pulse amplitude, and aperiodic X-ray luminosity, \cite{Yang_2017} also approximated $\tn{d}P/\tn{d}t$ from a linear fit of significant (at the $99\%$ level) $P(t_n)$ detections versus $t_n$ for the $52$ HMXBs in their library. Retaining timing points with statistical significance $\geq 99\%$ is standard in X-ray timing analysis. They are identified by searching each light curve for pulsations and calculating the Lomb-Scargle power for each independent frequency using Equation (2) in \cite{Yang_2017}. The statistically significant measurements are overplotted as shaded columns in Figure \ref{fig:ObsDistribution} and labeled as ``Detections''. Once retrieved, \cite{Yang_2017} characterized the rotational state of each star in terms of $\tn{d}P/\tn{d}t$ divided by its measurement error, denoted by $\epsilon$ in Table 3 in \cite{Yang_2017}. They classified 22 stars in rotational disequilibrium, with 12 stars spinning up $(\epsilon <-1.5)$, and 10 spinning down $(\epsilon > 1.5)$. The remaining 30 stars are in rotational equilibrium $(-1.5 \leq \epsilon \leq 1.5)$. The spin distribution is visualized in Figure \ref{fig:SpinDistribution} using a $P$--$|\dot{P}|$ diagram. Stars classified as spinning up, down, and in equilibrium are plotted as red, blue, and green points respectively.  It is possible in principle that some objects experience one or more torque reversals somewhere between the first ($t_1$) and last ($t_N$) samples. However, upon inspecting visually the 52 $P(t_n)$ time series, we do not observe strong evidence for torque reversals; if they are present, they are masked by the accretion-driven spin fluctuations. For simplicity, therefore, we run the nonlinear Kalman filter on the full time series for each pulsar, assuming implicitly that each pulsar is characterized by a single set of asymptotic ensemble-averaged parameters $(\bar{\Omega},\bar{Q},\bar{S},\bar{\eta})$ for $t_1 \leq t \leq t_N$, where the number of samples $N$ per pulsar is listed in the eighth column of Table 2 in \cite{Yang_2017} and summarized in Figure \ref{fig:ObsDistribution} above.

\section{Magnetocentrifugal accretion parameters}\label{Sec:MagPE}

The parameter estimation framework returns a suite of data products for each star, namely the posteriors of the six fundamental model parameters $\bm{\Theta} = (\beta_1, \beta_2, \gamma_Q, \gamma_S, \sigma_{QQ}, \sigma_{SS})$ and the four inferred parameters $(\bar{Q}, \bar{S}, \bar{\eta}, \mu)$, the Kalman filter state estimates $\bm{X}_n = [\Omega(t_n),Q(t_n),S(t_n)]$ as functions of $t_n$, as well as auto- and cross-correlation coefficients involving the accretion rate, the Maxwell stress at the disk-magnetosphere boundary, and the RXTE PCA measurements $P(t)$ and $L(t)$. We inspect the output for each star and retain results which satisfy the following acceptance criteria: (i) the one-dimensional posteriors for $\beta_1$ and $\beta_2$ are unambiguously recovered, i.e.\ they are unimodal and there is no evidence of railing against the limits of the prior; and (ii) the posterior median of the radiative efficiency satisfies $0 < \bar{\eta} < 1$. Out of the 52 SMC HMXBs, 24 satisfy both (i) and (ii). We will revisit the 28 HMXBs that do not satisfy (i) and (ii) in a future manuscript, after data volumes with $ > 10^3$ samples become available. In this section we summarise the nested sampler output for the model parameters $\bm{\Theta} = (\beta_1, \beta_2, \gamma_Q, \gamma_S, \sigma_{QQ}, \sigma_{SS})$ for the 24 accepted objects. To orient the reader, the data products available for a single, exemplary X-ray pulsar, namely SXP $4.78$, are presented in Appendix \ref{App:WorkedExample}. We refer to the data products in Appendix \ref{App:WorkedExample} throughout the remainder of the paper when discussing results for the SMC HMXB population. 

The acceptance criteria discussed above are deliberately conservative and motivated as follows. The first criterion, i.e.\ unimodal one-dimensional $\beta_1$ and $\beta_2$ posteriors, is adopted in lieu of setting upper limits on the parameters $\beta_1$ and $\beta_2$, e.g.\ by accepting posteriors that are flat or rail against the prior limits.  This is the first time $\beta_1$ and $\beta_2$ have been measured in practice, so we adopt broad, uninformative priors as a first pass at the problem and elect not to report upper limits on $\beta_1$ and $\beta_2$ until a more refined picture of their distribution is revealed by future X-ray timing experiments. Similarly, in the paper we report median posterior values as point estimates for the inferred parameters and hence we adopt the second criterion, i.e.\ that the inferred radiative efficiency satisfies $0 < \bar{\eta} < 1$, on physical grounds.

\begin{figure}
\centering{
\includegraphics[width=1\textwidth, keepaspectratio]{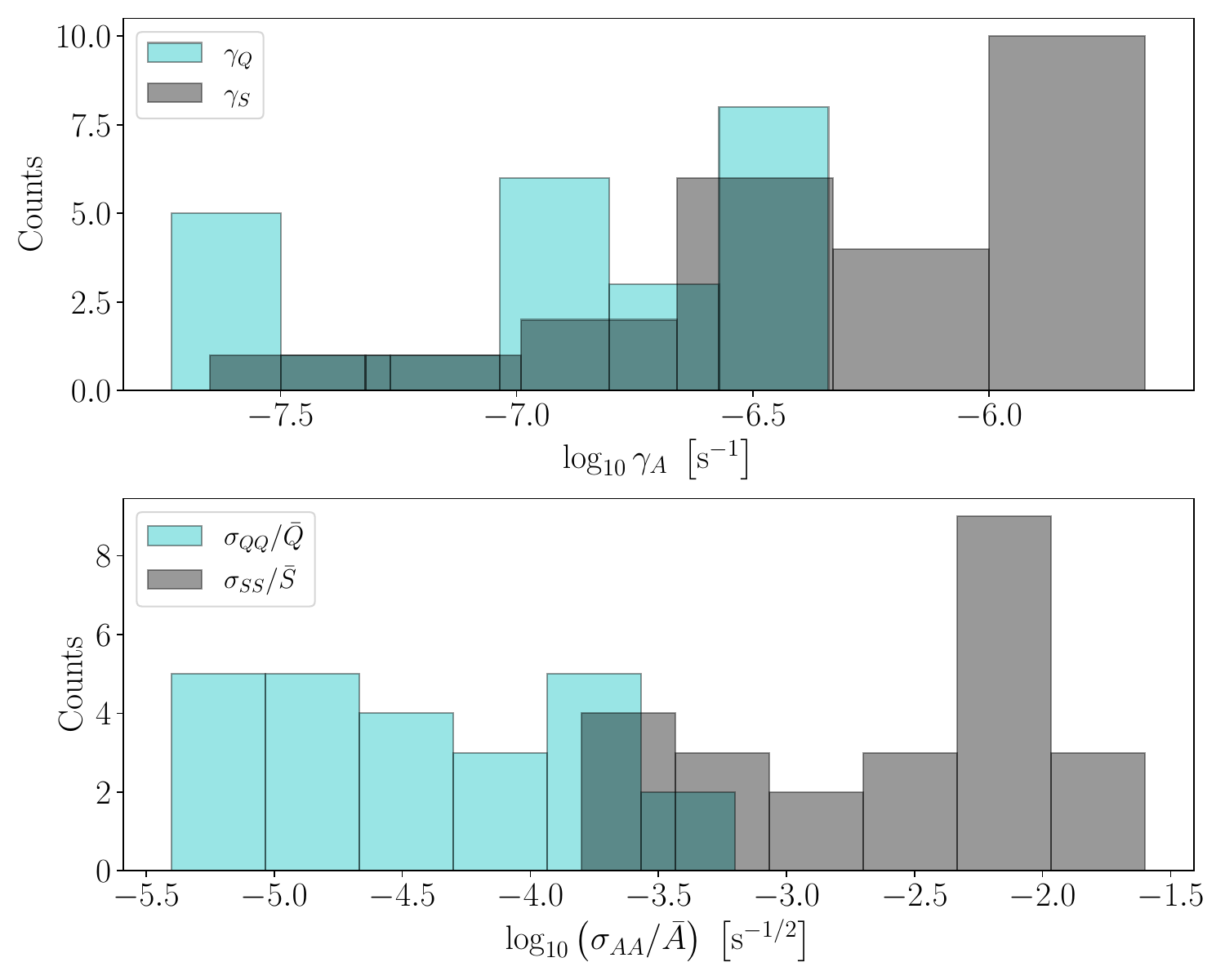}}
        \caption{Fluctuation parameters inferred by the nested sampler and Kalman filter from the RXTE PCA $P(t_n)$ and $L(t_n)$ time series for 24 SMC HMXBs. The top and bottom panels display histograms of posterior medians of $\gamma_A$ (units: $\rm{s}^{-1}$) and $\sigma_{AA}/\bar{A}$ (units: $\rm{s}^{-1/2}$) respectively, with $A=Q \, (\rm{cyan})$ and $S \, (\rm{gray})$.}
    \label{fig:ParamDistribution}
\end{figure}

The fluctuation parameters $(\gamma_A, \sigma_{AA}/\bar{A})$, with $A \in \{Q, S\}$, returned by the nested sampler and Kalman filter are summarized in Figure \ref{fig:ParamDistribution}. The top and bottom panels contain population histograms of the mean-reversion time scales, $\gamma_Q$ and  $\gamma_S$, and the normalized rms noise amplitudes, $\sigma_{QQ}/\bar{Q}$ and $\sigma_{SS}/\bar{S}$, respectively. In each panel, the reported values refer to the one-dimensional posterior median, a representative example of which is presented in Figure \ref{fig:AppWECornerPlot} in Appendix \ref{App:WorkedExample} for SXP $4.78$.

The results presented in Figure \ref{fig:ParamDistribution} have two key features. First, the characteristic time scale of mean reversion in the top panel satisfies $-8 \lesssim \log_{10} \left[\gamma_A/( 1 \, \rm{s}^{-1}) \right] \lesssim -6$ for $A \in \{Q, S \}$, which is broadly consistent with Figure 3 in \cite{Melatos_2022}, the results presented in Section 4.5 in \cite{OLeary_2023}, and the Lomb-Scargle PSD computed from the light curves of other accretion-powered systems such as Scorpius X$-$1 and 2S 1417$-$624, which roll over at $\sim 10^{-7} \, \rm{s^{-1}}$ \citep{Mukherjee_2018,Serim_2022}. Interestingly from a physical standpoint, the gray histograms in the top panel of Figure \ref{fig:ParamDistribution} lie systematically to the right of the cyan histograms, implying that the characteristic time scale of mean-reversion is shorter for $S$ fluctuations than for $Q$ fluctuations in the sample of SMC HMXBs studied here. Second, the inferred accretion rate fluctuations satisfy $\sigma_{QQ} \approx 0.2 \bar{Q}\gamma_Q^{1/2}$, which is broadly consistent with observations \citep{Serim_2022}; see Section 4.1 and Footnote 10 in \cite{Melatos_2022} for further details. Independent estimates of the Maxwell stress fluctuations $\sigma_{SS}/\bar{S}$ are not available for comparison in the literature at the population level for real objects in the SMC; they are measured in this paper for the first time (gray columns in the second panel of Figure \ref{fig:ParamDistribution}). 

\begin{figure}
\centering{
\includegraphics[width=1\textwidth, keepaspectratio]{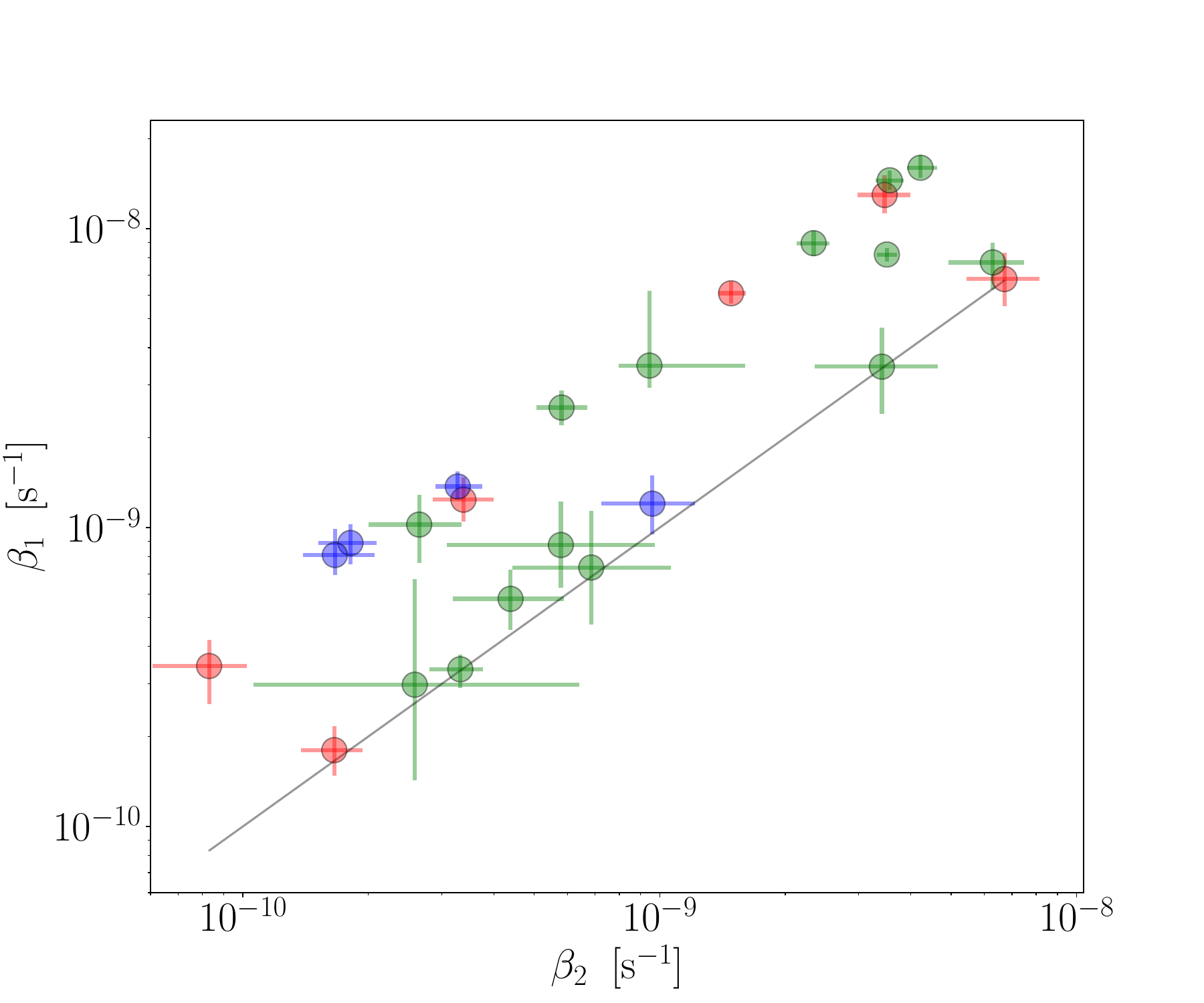}}
        \caption{Magnetocentrifugal parameters $\beta_1$ (units: $\rm{s^{-1}}$) and $\beta_2$ (units: $\rm{s^{-1}}$) inferred by the nested sampler and Kalman filter from the RXTE PCA $P(t_n)$ and $L(t_n)$ time series for 24 HMXBs in the SMC. Stars classified as spinning up, down, and near rotational equilibrium are plotted as red, blue, and green points, respectively. The vertical and horizontal error bars correspond to the 68\% credible interval associated with the one-dimensional posteriors of $\beta_1$ and $\beta_2$, respectively. The gray, diagonal line indicates $\beta_1=\beta_2$. The vertical and horizontal axes are plotted on $\log_{10}$ scales.}
    \label{fig:BetaDistribution}
\end{figure}

In Figure \ref{fig:BetaDistribution} we plot $\beta_1$  as a function of $\beta_2$. The plotted points and associated error bars correspond to the one-dimensional posterior median and the 68\% credible interval, respectively. Stars classified as spinning up, down, and in rotational equilibrium are plotted as red, blue, and green points, respectively. For systems near rotational equilibrium, we expect $\beta_1 \approx \beta_2$ upon substituting $\bar{\Omega} = 2^{3/2} \pi^{3/5} (GM)^{1/5} \bar{Q}^{-3/5}\bar{S}^{3/5}$ in the denominator of Equation (\ref{eq:beta1}) above where the latter prescription for $\bar{\Omega}$ is valid for equilibrium only. Therefore it is encouraging that seven out of the 14 green points lie on the gray diagonal line $\beta_1=\beta_2$, and the remaining seven lie near the line, with $0.4 \, {\rm dex} \lesssim | \beta_2 - \beta_1 | \lesssim 0.6 \, {\rm dex}$. We also infer $\beta_1 \approx \beta_2$ for two stars classified as spinning up, namely SXP 4.78 and SXP 11.9 (red points), and one star classified as spinning down, namely SXP 138 (blue point). It is interesting astrophysically that we find $\beta_1 \gtrsim \beta_2$ in Figure \ref{fig:BetaDistribution}. Using Equations (\ref{eq:beta1}), (\ref{eq:beta2}), and (\ref{eq:MuEquation}), one can deduce an inequality between the magnetic moment $\mu$ and the state variables $\bar{\Omega}, \bar{Q},$ and $\bar{S}$:
\begin{equation}
    \mu \gtrsim 0.014 (GM)^{2/5} \,\bar{\Omega}\,\bar{Q}^{9/5}\, \bar{S}^{-13/10}.
\end{equation}
The conversion of $\beta_1$ and $\beta_2$ into posterior estimates of $\mu$ and $\bar{\eta}$ is discussed in Sections \ref{Sec:MagMoms} and \ref{Sec:eta} respectively. 

The results presented in Figure \ref{fig:BetaDistribution} hint at the existence of two parallel trends in the $\beta_1$-$\beta_2$ plane. At the time of writing, it is unclear whether or not the trends are astrophysical, as Figure \ref{fig:BetaDistribution} contains only 24 data points, and some of the error bars are substantial. In order to test for a numerical artefact arising from the parameter estimation framework, we do the following analyses. (i) We draw 30 samples randomly from the six-dimensional posterior distribution for each of the 24 objects, and plot the $\beta_1$ samples as a function of the $\beta_2$ samples for each object. Interestingly, we observe that the two parallel trends visible in Figure \ref{fig:BetaDistribution} are still present. (ii) We run the unscented Kalman filter on the linear, magnetocentrifugal state-space model given by Equations (13)--(16) in \cite{Melatos_2022}. The output of the unscented Kalman filter coincides with the output of a linear Kalman filter, as expected.

One possible astrophysical interpretation of the parallel trends in Figure \ref{fig:BetaDistribution} is as follows. Global, three-dimensional magnetohydrodynamic numerical simulations studying Rayleigh-Taylor instabilities at the disk-magnetosphere boundary of rotating magnetized stars \citep{Romanova_2014,Romanova_2015,Blinova_2016} reveal that accretion occurs in three regimes, governed by the time-dependent fastness parameter $[R_{\rm m}(t)/R_{\rm c}(t)]^{3/2}$. In a time-averaged sense, it is possible that the results in Figure \ref{fig:BetaDistribution} are connected with two of the three accretion regimes identified by \cite{Blinova_2016}. We discuss this possibility briefly in Section \ref{SubSec:MagMomentObs} and will explore the issue in detail in a forthcoming paper.

The subsample of spin disequilibrium SMC HMXBs with $\beta_1 \approx \beta_2$, i.e.\ SXP 4.78, SXP 11.9, and SXP 138, are interesting from an observational standpoint for the following reason. \cite{Yang_2017} measured $\tn{d}P/\tn{d}t$ using $\leq 11$ significant timing points per object, e.g.\ $\tn{d}P/\tn{d}t$ is approximated for SXP 11.9 using three significant (at the 99\% level) pulse period measurements over $\sim 16$ years. Given the limited volume of data available for SXP 4.78, SXP 11.9, and SXP 138, it is challenging to measure $\tn{d}P/\tn{d}t$ with a high level of confidence using time-averaged techniques. Accordingly, the Kalman filter parameter estimation scheme may serve as an important practical tool in the future for testing whether an X-ray transient is in magnetocentrifugal equilibrium or not.

The uncertainty on each median counted in Figures \ref{fig:ParamDistribution} and \ref{fig:BetaDistribution} can be estimated by the full width half maximum (FWHM) of the corresponding one-dimensional posterior, e.g.\ the FWHM in Figure \ref{fig:AppWECornerPlot} ranges from a minimum of $\approx 0.12 \; \rm{dex}$ for $\sigma_{QQ}$ to a maximum of $\approx 0.55  \; \rm{dex}$ for $\gamma_S$. The dispersion is modest in general. For example, the FWHM for the 24 accepted objects ranges from $\approx 0.30 \; \rm{dex}$ for $\beta_1$ and $\beta_2$ to $\approx 0.85 \; \rm{dex}$ for $\gamma_Q$, which agrees approximately with the Monte Carlo validation tests performed in \cite{Melatos_2022} and the results presented in Section 4.5 in \cite{OLeary_2023} for the X-ray transient SXP 18.3. The FWHM for individual objects is smaller typically than the population-wide range spanned by the horizontal axes in Figure \ref{fig:ParamDistribution}.

We remind the reader of two points. (i) The chief goal of the present paper is to infer $\mu$ for a subsample of SMC HMXBs without any attendant assumptions about linearity or rotational equilibrium, as in \cite{Melatos_2022} and \cite{OLeary_2023}. (ii) The framework presented in this paper is an extension of the linear framework developed by \cite{Melatos_2022} to the nonlinear regime and rotational disequilibrium using real observational data. With respect to point (i) above, the additional data products in Sections \ref{Sec:MagPE}, \ref{Sec:eta}, and \ref{Sec:Corr}, e.g.\ new estimates of the Pearson correlation coefficients for a suite of variable pairs for 24 SMC HMXBs, are presented here in preliminary form to highlight the new information available with a Kalman filter analyses of X-ray timing data, over and above standard, time-averaged techniques. The quantitative details will be explored fully in future analyses. With respect to point (ii) above, this is the first time the Kalman filter has been applied to rotational disequilibrium using real observational data, so we elect as a first pass not to generalize the parameter estimation framework proposed by \cite{Melatos_2022}, i.e.\ we infer $\bar{Q}, \bar{S}, \bar{\eta},$ and $\mu$ using nonlinear combinations of the inferred $\beta_1$ and $\beta_2$ parameters, discussed in detail in Section \ref{Sec:UKF}. Other sampling strategies exist, e.g.\ one may elect to sample $\bar{\eta}$ or $\bar{Q}$ directly using physically motivated, restrictive priors on $\bar{\eta}$ and $\bar{Q}$, or work with implicit priors for $\bar{\eta}$ or $\bar{Q}$. Such opportunities are not explored in this paper but are mentioned to support future Kalman filter applications.

\section{Magnetic moments}\label{Sec:MagMoms}

\begin{figure}
\centering{
\includegraphics[width=1\textwidth, keepaspectratio]{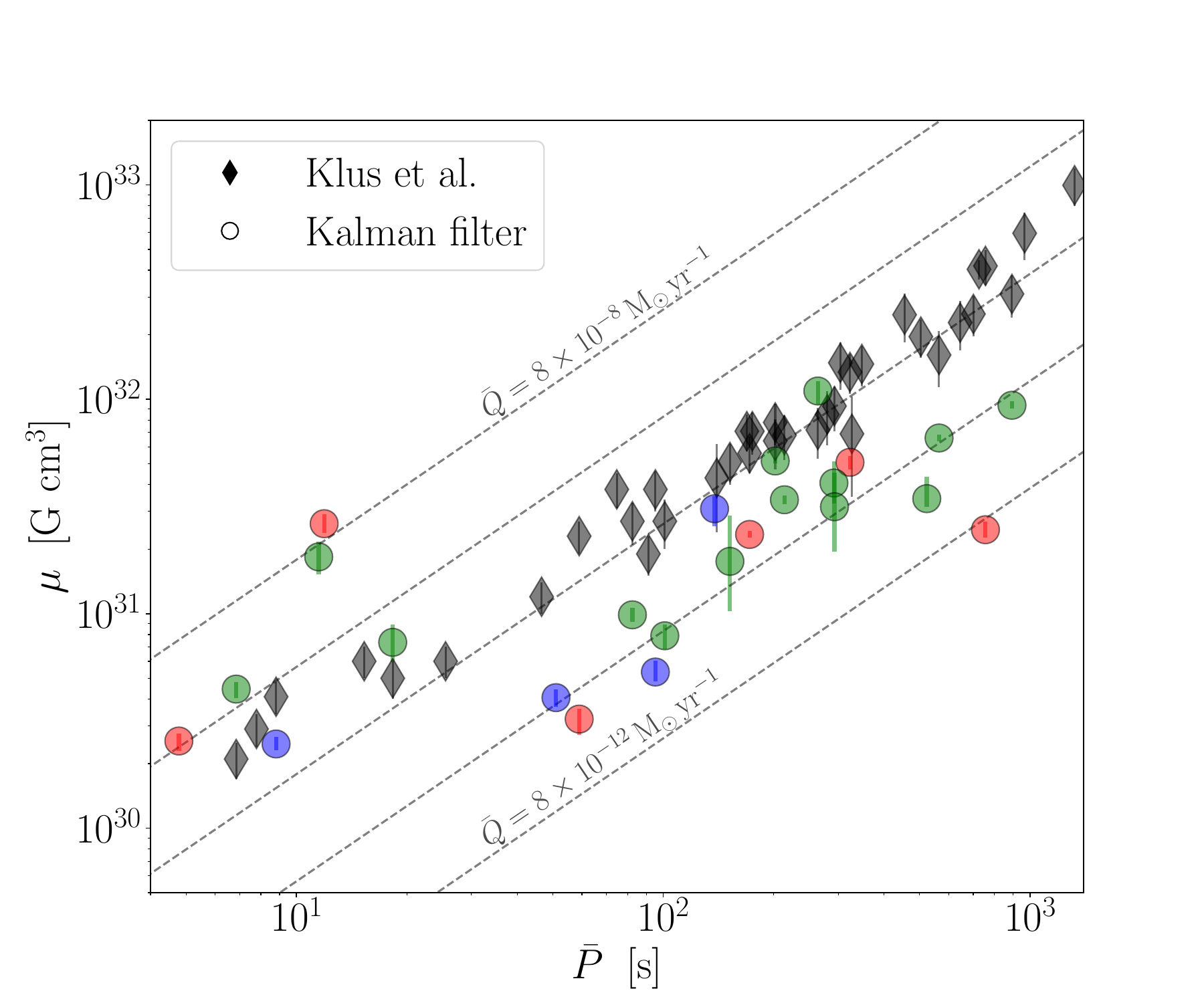}}
        \caption{Magnetic dipole moments $\mu$ (units: $\rm{G \, cm^3}$) versus time-averaged pulse period $\bar{P}$ (units: $\rm{s}$) for the 24 accepted SMC HMXBs. The plotted points and associated error bars correspond to the one-dimensional posterior median and the 68\% credible interval inferred by the nested sampler from the RXTE PCA $P(t_n)$ and $L(t_n)$ time series, respectively. Stars classified as spinning up, down, and in rotational equilibrium are plotted using red, blue, and green points, respectively. Magnetic dipole moments and associated uncertainties  inferred from Table 3 in \cite{Klus_2014} (for an overlapping but different set of objects) are overplotted as gray diamonds. The five gray, dashed lines are representative $\mu$ values, calculated using the time-averaged magnetocentrifugal spin-up line, i.e.\ Equation (6) in \cite{OLeary_2023}, for $8\, \times \, 10^{-12} \leq \bar{Q}/(1 \, \rm{M_\odot \, yr^{-1}}) \leq 8 \, \times \, 10^{-8}$, spaced by $1 \, \rm{dex}$. The vertical and horizontal axes are plotted on $\log_{10}$ scales.}
    \label{fig:MuDistribution}
\end{figure}

The magnetic dipole moments returned by the nested sampler and Kalman filter for the 24 accepted SMC HMXBs are presented in Figure \ref{fig:MuDistribution} as a function of $\bar{P}$. The analysis yields 24 independent $\mu$ estimates, with six stars classified as spinning up (red points), four as spinning down (blue points), and 14 as being near rotational equilibrium (green points). The plotted points correspond to the posterior median, and the error bars correspond to the  $68\%$ credible interval. The five gray, dashed lines denote representative $\mu$ values, calculated using the time-averaged magnetocentrifugal spin-up line relating $\mu$ to $\bar{\Omega}$ and $\bar{Q}$, i.e.\ Equation (6) in \cite{OLeary_2023}, assuming fixed mass accretion rates, with $8 \times 10^{-12} \leq \bar{Q}/(1\, \rm{M_\odot \, yr^{-1}}) \leq 8 \times 10^{-8}$. The inferred $\mu$ estimates from Table 3 in \cite{Klus_2014} are overplotted as gray diamonds for comparison. For stars classified as spinning up (red points), $\mu$ ranges from a minimum of $2.5^{+0.2}_{-0.3} \times 10^{30} \, \rm{G \ cm^3}$ for SXP 4.78 to a maximum of $5.1^{+0.4}_{-0.4} \times 10^{31} \, \rm{G \ cm^3}$ for SXP 323. For stars classified as spinning down (blue points), $\mu$ ranges from a minimum of $2.5^{+0.2}_{-0.2} \times 10^{30} \, \rm{G \ cm^3}$ for SXP 8.80 to a maximum of $3.0^{+0.4}_{-0.5} \times 10^{31} \, \rm{G \ cm^3}$ for SXP 138. For stars in rotational equilibrium (green points), $\mu$ ranges from a minimum of $4.5^{+0.3}_{-0.4} \times 10^{30} \, \rm{G \ cm^3}$ for SXP 6.85 to a maximum of $1.1^{+0.11}_{-0.15} \times 10^{32} \, \rm{G \ cm^3}$ for SXP 264.

In the absence of independent local magnetic field strength measurements \citep{Staubert_2019}, i.e.\ by identifying resonant cyclotron features in the X-ray spectra of SMC HMXBs, it is challenging to verify the accuracy with which the Kalman filter and nested sampler estimate $\mu$ at a population level. Yet the results in Figure \ref{fig:MuDistribution} are interesting for the following two reasons. First, the Kalman filter and nested sampler return consistent answers, whether one assumes magnetocentrifugal equilibrium or not \citep{Melatos_2022}. For example, \cite{OLeary_2023} obtained $\mu = 8.0^{+1.3}_{-1.2} \times 10^{30} \rm{\, G \, cm^3}$ assuming magnetocentrifugal equilibrium for the SMC X-ray transient SXP 18.3 (with $N = 854$ samples), while the present analysis returns $\mu = 7.4^{+1.5}_{-1.4} \times 10^{30} \rm{\, G \, cm^3}$ without assuming magnetocentrifugal equilibrium. Second, the analysis in this paper yields $\mu$ values which are smaller than those obtained using time-averaged statistics for $\bar{P} \gtrsim 30 \, \rm{s}$, e.g.\ the difference between $\mu$ inferred for SXP 756 by \cite{Klus_2014} and the present analysis is $\approx 1.2 \, \rm{\, dex}$. The complete set of  objects analyzed by \cite{Klus_2014} and in the present paper are reported in Table \ref{tab:KlusMuOverlap}.

\begin{table}
\centering
\setlength{\tabcolsep}{5pt}
\hspace*{-2.0cm}
 \begin{tabular}{c c c c} 
 \hline
  Pulsar & $\mu/(\rm{10^{31}\, G \, cm^3})$ (Kalman filter)  & $\mu/(\rm{10^{31}\, G \, cm^3})$ \citep{Klus_2014} & Difference [\rm{dex}]  \\ [1ex] 
  \hline
 \hline 
SXP 6.85 & $0.45^{+0.03}_{-0.04}$ & $0.21^{+0.04}_{-0.04}$ & $0.33$ \\ [1ex]
SXP 8.80 & $0.25^{+0.02}_{-0.02}$ & $0.41^{+0.07}_{-0.07}$ & $-0.21$ \\ [1ex]
SXP 18.3 & $0.74^{+0.15}_{-0.14}$ & $0.50^{+0.1}_{-0.1}$ & $0.17$ \\ [1ex]
SXP 59.0 & $0.32^{+0.04}_{-0.05}$ & $2.3^{+0.4}_{-0.4}$ & $-0.86$ \\ [1ex]
SXP 82.4 & $1.0^{+0.07}_{-0.07}$ & $2.7^{+0.6}_{-0.6}$ & $-0.43$ \\ [1ex]
SXP 95.2 & $0.53^{+0.07}_{-0.05}$ & $3.8^{+0.8}_{-0.8}$ & $-0.86$ \\[1ex]
SXP 101 & $0.79^{+0.10}_{-0.11}$ & $2.7^{+0.7}_{-0.7}$ & $-0.53$ \\ [1ex]
SXP 152 & $1.8^{+1.10}_{-0.73}$ & $5.1^{+1.1}_{-1.1}$ & $-0.45$ \\ [1ex]
SXP 172 & $2.3^{+0.09}_{-0.08}$ & $5.6^{+1.1}_{-1.1}$ & $-0.39$ \\ [1ex]
SXP 202A & $5.2^{+0.19}_{-0.15}$ & $7.8^{+1.7}_{-1.7}$ & $-0.18$ \\ [1ex]
SXP 214 & $3.4^{+0.15}_{-0.16}$ & $6.8^{+1.6}_{-1.6}$ & $-0.30$ \\ [1ex]
SXP 264 & $11.0^{+1.1}_{-1.5}$ & $7.2^{+1.9}_{-1.9}$ & $0.18$ \\ [1ex]
SXP 293 & $3.2^{+1.4}_{-1.2}$ & $9.3^{+2.2}_{-2.2}$ & $-0.46$ \\ [1ex]
SXP 323 & $5.1^{+0.40}_{-0.40}$ & $13.4^{+2.8}_{-2.8}$ & $-0.42$ \\ [1ex]
SXP 565 & $6.6^{+0.26}_{-0.25}$ & $16.1^{+4.7}_{-4.7}$ & $-0.39$ \\ [1ex]
SXP 756 & $2.5^{+0.22}_{-0.20}$ & $41.9^{+7.9}_{-7.9}$ & $-1.20$ \\ [1ex]
SXP 893 & $9.4^{+0.43}_{-0.35}$ & $31.0^{+6.9}_{-6.9}$ & $-0.52$ \\ [1ex]
\hline
 \end{tabular}
 \caption{Subsample of dual $\mu$ estimates for the SMC HMXBs analyzed by \cite{Klus_2014} and in the present paper. We report the Kalman filter $\mu$ estimates in the second column, the inferred $\mu$ values from Table 3 in \cite{Klus_2014} in the third column, and the difference between the estimates in the fourth column. A positive value in the fourth column indicates that the value reported in the present paper exceeds the value reported in \cite{Klus_2014}.}
 \label{tab:KlusMuOverlap}
\end{table}

The systematic difference arises due to the following subtle astrophysical reason. \cite{Klus_2014} did not estimate $\bar{Q}$ and $\bar{\eta}$ independently; they assumed $\bar{Q} \propto \langle L(t) \rangle$ and $\bar{\eta}=1$. The aperiodic X-ray luminosities reported in Table 1 in \cite{Klus_2014} exhibit little variation for $\bar{P} \gtrsim 10^2 \, \rm{s}$, with $0.2 \lesssim \bar{L}/(10^{37}\,\rm{erg \, s^{-1}}) \lesssim  0.9$, so the inferred magnetic moments approximately satisfy $\mu \propto \bar{P}^{7/6}$ \citep{Klus_2014}. The scaling $\mu \propto \bar{P}^{7/6}$ is visible in Figure \ref{fig:MuDistribution} where the gray diamonds match closely the gray dashed line, calculated using the time-averaged magnetocentrifugal spin-up line relating $\mu$ to $\bar{\Omega}$ and $\bar{Q}$, i.e.\ Equation (6) in \cite{OLeary_2023}, assuming $\bar{Q} = 8 \, \times 10^{-10} \, \rm{M_\odot \, yr^{-1}}$. In the present analysis, however, the parameters $(\bar{Q}, \bar{S}, \bar{\eta})$ are estimated independently, implying $\mu \propto \bar{Q}^{6/5}\bar{S}^{-7/10}$. 

In Figure \ref{fig:MuDistribution}, we do not report new $\mu$ estimates for SXP 15.3 \citep{Maitra_2018} or SXP 2.37 \citep{jaisawal_2016}, whose $\mu$ values are measured directly using cyclotron resonant scattering features, for the following astrophysical reasons. In SXP 15.3, the nested sampler returns $\bar{\eta} > 1$. A radiative efficiency above unity is unphysical, but is permitted mathematically by Equation (\ref{eq:EtaConstant}). In SXP 2.37, the nested sampler returns an accretion rate well above the Eddington limit, with $|Q_1(t)|\bar{Q} \sim 1\times 10^{-5} \rm{\, M_{\odot} \, yr^{-1}}$, and is vetoed accordingly. Detailed calibration of the parameter estimation scheme using $P(t_n)$ and $L(t_n)$ measurements of Galactic X-ray pulsars, more of whose local magnetic field strengths are measured via cyclotron absorption lines, will be the topic of a future paper. 

\section{Radiative efficiency}\label{Sec:eta}

In the canonical picture of magnetocentrifugal accretion, the radiative efficiency is degenerate with the mass accretion rate; that is, $\bar{Q}$ and $\bar{\eta}$ appear combined as a product in Equation (\ref{eq:AveragedLstate}). To circumvent this issue, it is standard practice to assume a specific value for $\bar{\eta}$. Order-of-magnitude analyses generally assume $\bar{\eta
} = 1$, i.e.\ $100\%$ of the gravitational potential energy of material falling onto the star is converted to heat and hence X-rays. Other studies assume $0.1 \lesssim \bar{\eta} \lesssim 0.2$ \citep{klochkov_2009,doroshenko_2010} on theoretical grounds \citep{Frank_2002, Longair_2010}. Neither assumption has been verified by measuring $\bar{\eta}$ independently, a challenging task. 

\begin{figure}
\centering{
\includegraphics[width=1\textwidth, keepaspectratio]{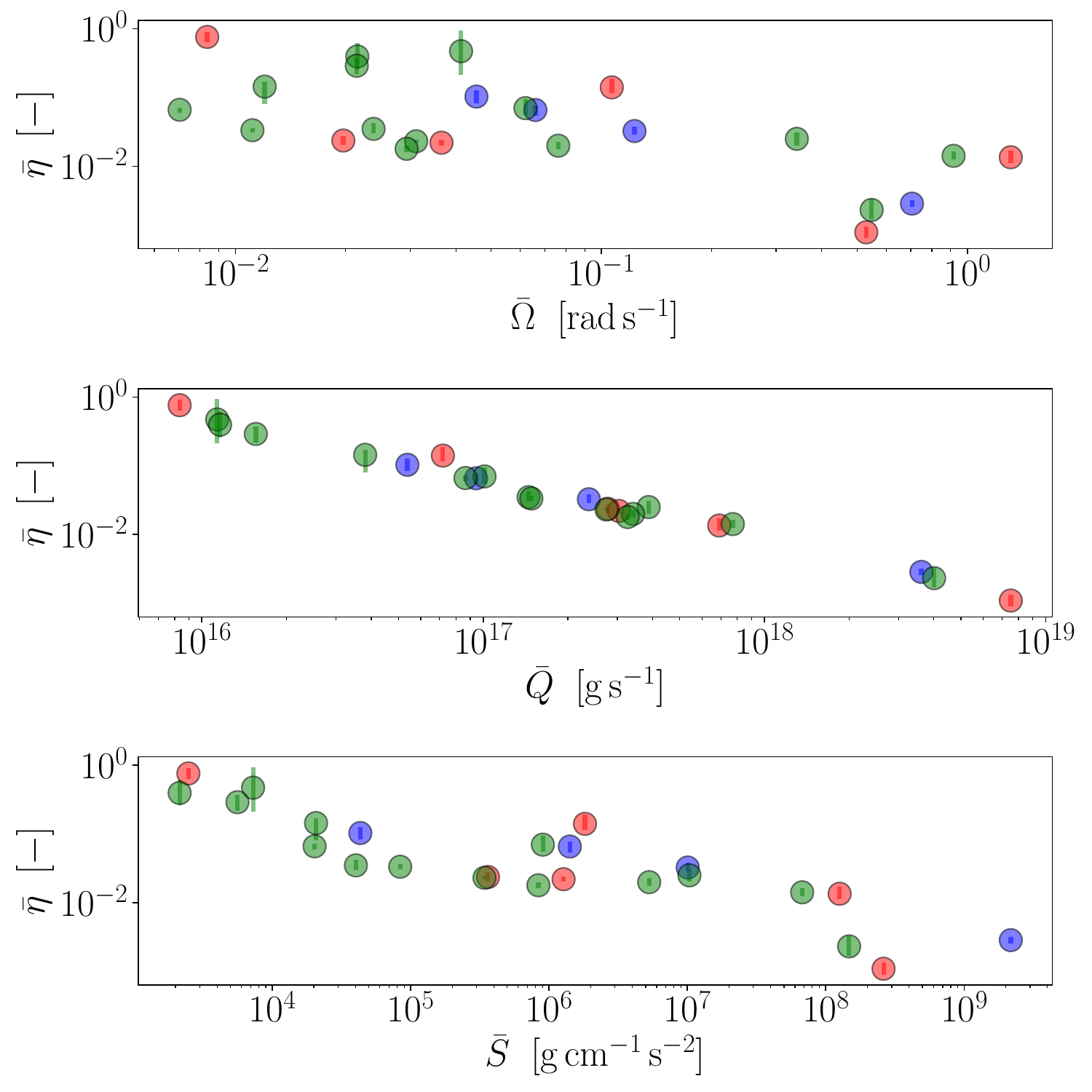}}
        \caption{Radiative efficiency $\bar{\eta}$ (units: dimensionless) inferred by the nested sampler from the RXTE PCA $P(t_n)$ and $L(t_n)$ time series for 24 SMC HMXBs. The radiative efficiency $\bar{\eta}$ is plotted as a function of the angular velocity $\bar{\Omega}$ (units: $\rm{rad \,\, s^{-1}}$; top panel), the mass accretion rate $\bar{Q}$ (units: $\rm{g \, s^{-1}}$; middle panel), and the scalar Maxwell stress $\bar{S}$ (units: $\rm{g \, cm^{-1} \, s^{-2}}$; bottom panel). The plotted points refer to the one-dimensional posterior median, the error bars correspond to the 68\% credible interval, and stars classified as spinning up, down, and in rotational equilibrium are plotted using red, blue, and green points, respectively. The vertical and horizontal axes are plotted on $\log_{10}$ scales. }
    \label{fig:EtaDistribution}
\end{figure}

We report the nested sampler estimates of $\bar{\eta}$ as a function of the asymptotic ensemble-averaged parameters $\bar{\Omega}, \bar{Q},$ and $\bar{S}$   in the top, middle, and bottom panels of Figure \ref{fig:EtaDistribution}, respectively. The plotted points refer to the one-dimensional posterior median, the error bars correspond to the $68\%$ credible interval, and sources classified as spinning up, down, or near equilibrium are plotted as red, blue, and green circles, respectively. The estimated $\bar{\eta}$ values range from $0.1\%$ to $76\%$ for stars classified as spinning up (red points), from $0.2\%$ to $10\%$ for stars classified as spinning down (blue points), and from $0.2\%$ to $47\%$ for stars in equilibrium (green points). The median and mean of the $\bar{\eta}$ distribution are $3\%$ and $12\%$ respectively. Out of the 24 accepted SMC HMXBs, 14 objects satisfy $0.01 \lesssim \bar{\eta} \lesssim 0.09$, three objects lie within the traditional (but approximate) theoretical range $0.1 \lesssim \bar{\eta} \lesssim 0.2$, four objects satisfy $0.3 \lesssim \bar{\eta} \lesssim 0.8$, and the remaining three objects satisfy $0.001 \lesssim \bar{\eta} \lesssim 0.003$. In some cases, the analysis returns point estimates of $\bar{\eta}$ above unity, as flagged in Section \ref{Sec:MagPE}. We elect not to report new results for these objects, but note that some segments of the inferred posterior are physical and some others are not, e.g.\ the inferred posteriors cover  $0.48 < \bar{\eta} < 3.9$ for SXP 7.92, $0.99 < \bar{\eta} < 4.2$ for SXP 8.02, and $0.48 < \bar{\eta}<2.45$ for SXP 15.3, of which the segments $\bar{\eta} < 1$ are physical.

Figure \ref{fig:EtaDistribution} contains evidence of moderately significant anticorrelations in the second and third panels.  Linear fits of $\log_{10} \bar{\eta}$ versus $\log_{10} \bar{Q}$ and $\log_{10}\bar{S}$ reveal approximate power-law relationships of the form $\bar{\eta} \propto \bar{Q}^{-9/10}$ and $\bar{\eta} \propto \bar{S}^{-9/25}$, with associated coefficients of determination $R^2_{\bar{\eta},\bar{Q}} \approx 0.98$ and $ R^2_{\bar{\eta},\bar{S}} \approx 0.75$, respectively. Arguably there is some evidence for an anticorrelation $\bar{\eta} \propto \bar{\Omega}^{-7/10}$  in the top panel but it is marginal ($R^2_{\bar{\eta},\bar{\Omega}} = 0.45$) and is not examined further here. The $\bar{\eta}$ scalings in Figure \ref{fig:EtaDistribution} are examples of the new and astrophysically important information that stems from the time-resolved Kalman filter analysis and complements the output of time-averaged analyses. One possible physical interpretation follows. Three-dimensional hydrodynamical simulations \citep{Romanova_2015} reveal that the accretion physics in Section \ref{Sec:MeasuringMagMom} is supplemented by outflows, launched by several complex mechanisms, e.g.\ the accretion flow is redirected outwards along open magnetic field lines \citep{Matt_2005,Matt_2008,Dangelo_2017}; see Appendix C.4 in \cite{Melatos_2022} for a brief survey of accretion outflows in this context. It is plausible that outflows are stronger, when $\bar{Q}$ is high and pressure builds up rapidly at the disk-magnetosphere boundary \citep{Marino_2019}, so that $\bar{\eta}$ decreases as $\bar{Q}$ increases, as seen in the middle panel of Figure \ref{fig:EtaDistribution}. The scaling $\bar{\eta} \propto \bar{S}^{-9/25}$ is more challenging to interpret astrophysically. However, in a time-averaged sense, the magnetosphere compresses, when $Q(t)$ exceeds $\bar{Q}$, producing spikes in $S(t)$; that is, $\bar{S}$ increases as $\bar{\eta}$ decreases, as seen in the bottom panel of Figure \ref{fig:EtaDistribution}.

There are two practical subtleties that should be taken into account when interpreting the results in Figure \ref{fig:EtaDistribution}. First, in this section we apply the parameter estimation framework to all observations in Figure \ref{fig:ObsDistribution}, i.e.\ we do not restrict attention to significant detections only \citep{Yang_2017}. As a result, the sample means $\bar{L}$ calculated in this paper are generally (albeit marginally) smaller than those considered in (for example) \cite{Klus_2014}. Significant detections are recorded during outbursts, when $\bar{L}$ is higher; see the green points in the second panel of Figure 1 in \cite{OLeary_2023}. Accordingly, further analyses are required using $P(t)$ and $L(t)$ data from sources  which benefit from more significant detections than those analyzed here, e.g.\ 2A 1822$–$37, and XB
1916$–$053 \citep{Patruno_2017,Patruno_2021}. Second, it is tempting to ascribe the $\bar{\eta}$ versus $\bar{Q}$ trend to the following artificial fitting effect. Equation (A6) in \cite{Melatos_2022} implies $\bar{\eta} \propto \bar{Q}^{-1}$, provided that $R \bar{L}/[(GM)^{1/3}I \bar{\Omega}^{4/3}]$ is roughly constant across the population. It is unclear why $R \bar{L}/[(GM)^{1/3}I \bar{\Omega}^{4/3}]$ should be roughly constant, but one could imagine a selection effect being responsible. However, a selection effect argument is unlikely to hold because it would also imply $\bar{\eta} \propto \bar{S}^{-1}$ from Equation (A5) in \cite{Melatos_2022}, which contradicts the trend in the bottom panel of Figure \ref{fig:EtaDistribution}.

\section{Temporal correlations}\label{Sec:Corr}

The Kalman filter state estimates $\bm{X}_n = [\Omega(t_n), Q(t_n), S(t_n)]$ are generated using the posterior maximum likelihood estimate of $\bm{\Theta} = (\beta_1, \beta_2, \gamma_Q, \gamma_S, \sigma_{QQ}, \sigma_{SS})$ as optimal  point estimates. Examples of the inferred time series can be found in Figure \ref{fig:AppWEKFTracking} below, and Figures 1 in \cite{Melatos_2022} and \cite{OLeary_2023}. It is instructive physically to cross-correlate the state estimates $\bm{X}_n$ with the time series of the measurement variables $P(t)$ and $L(t)$. Among other goals, one can check whether traditional assumptions about the system's hidden behavior hold, e.g.\ that $Q(t)$ increases, when $L(t)$ increases. 

\begin{figure}
\centering{
\includegraphics[width=1\textwidth, keepaspectratio]{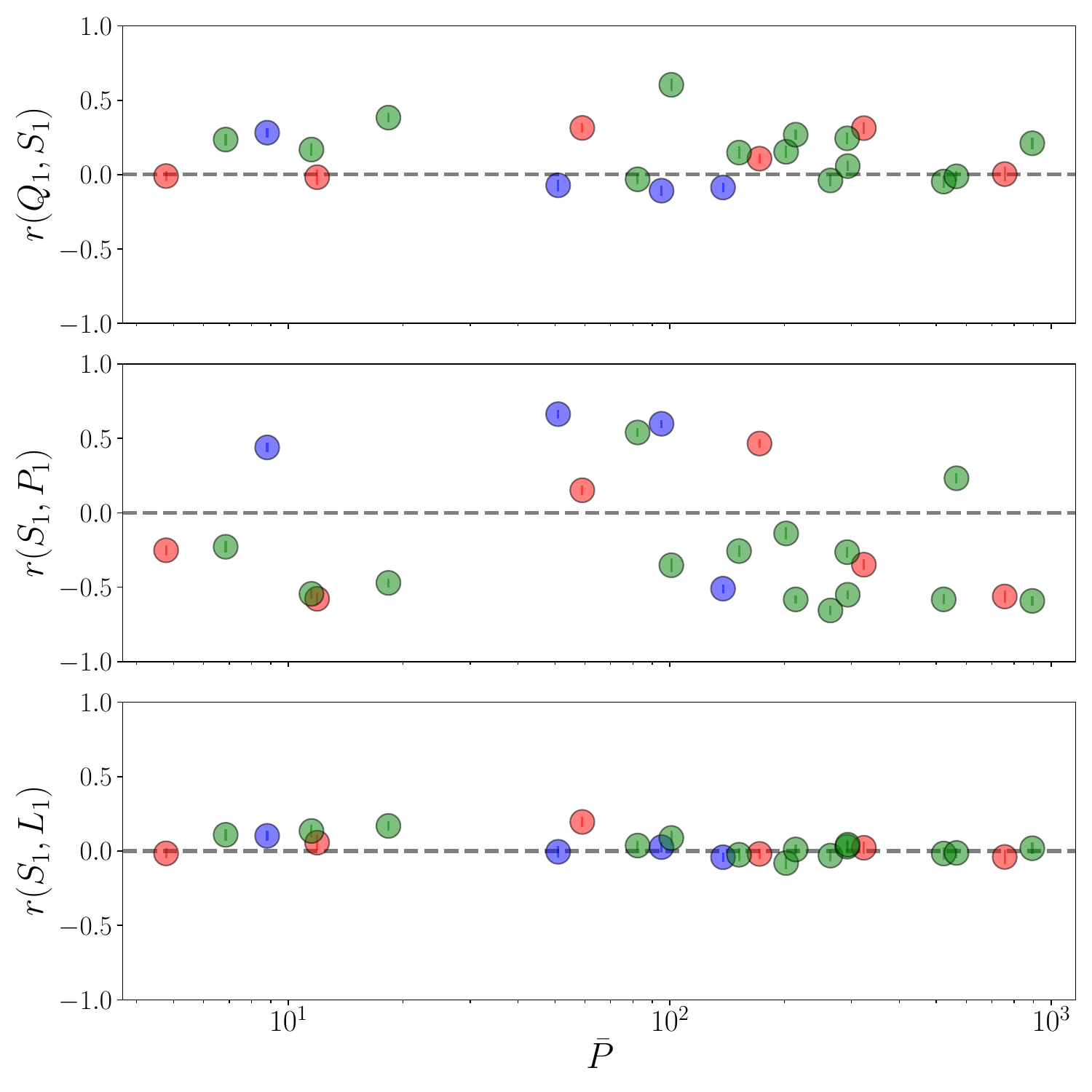}}
        \caption{Pearson correlation coefficients between the Maxwell stress at the disk-magnetosphere boundary and (i) the mass accretion rate $r(Q_1,S_1)$ (top panel); (ii) the pulse period $r(S_1,P_1)$ (middle panel); and (iii) the aperiodic X-ray luminosity $r(S_1, L_1)$ (bottom panel), versus the time-averaged pulse period $\bar{P}$.  The error bars $s_r$  correspond to the standard errors on $r$, which satisfy $0.03 \lesssim s_r \lesssim 0.05$. Stars classified as spinning up, down, and in rotational equilibrium are plotted using red, blue, and green points, respectively. The horizontal axes are plotted on $\log_{10}$ scales. }
    \label{fig:CorrDistribution}
\end{figure}

In Figure \ref{fig:CorrDistribution}, we report new estimates of the Pearson correlation coefficients $r$ and associated standard errors $s_r$ for a suite of variable pairs for the 24 SMC HMXBs analyzed in this paper. Specifically, we focus on cross-correlating the scalar Maxwell stress, which cannot be measured directly, with the (i) accretion rate, $r[Q_1(t),S_1(t)]$; (ii) pulse period, $r[S_1(t),P_1(t)]$;  and (iii) aperiodic X-ray luminosity, $r[S_1(t),L_1(t)]$.  Stars classified as spinning up, down, and near equilibrium are plotted as red, blue, and green points, respectively. The error bars in Figure \ref{fig:CorrDistribution} satisfy $0.03 \lesssim s_r \lesssim 0.05$.  

In the top panel of Figure \ref{fig:CorrDistribution}, we find that $Q_1(t_n)$ and $S_1(t_n)$ are uncorrelated for three stars classified as spinning up (red points), three stars classified as spinning down (blue points), and five stars classified as being near rotational equilibrium (green points), which  satisfy $ -0.10 \lesssim r[Q_1(t), S_1(t)] \lesssim -0.01$. For the remaining 13 stars, we find evidence of moderate positive correlations, which range from a minimum of $r[Q_1(t), S_1(t)]= 0.15 \pm 0.04$ for SXP 152 to a maximum of $r[Q_1(t),S_1(t)]= 0.60 \pm 0.04$ for SXP 101. A positive correlation makes sense physically: as $Q_1(t)$ increases, the accretion flow compresses the disk-magnetosphere boundary inwards, and the local magnetic field strength at the boundary increases, everything else being equal. Interestingly, though, everything else is not always equal; in 11 out of 24 objects, other stochastic dynamics mask the baseline compression phenomenon noted above.

In the middle panel of Figure \ref{fig:CorrDistribution}, seven systems exhibit positive correlations which range from a minimum of $r[S_1(t),P_1(t)] = 0.15 \pm 0.03$ for SXP 59.0 to a maximum of $r[S_1(t),P_1(t)] = 0.66 \pm 0.03$ for SXP 51.0. The remaining 17 stars exhibit evidence of moderate anticorrelations and satisfy $-0.59 \leq r[S_1(t),P_1(t)] \leq -0.14$. It is interesting astrophysically that 12 out of the 14 stars classified as being near rotational equilibrium (green points) are anticorrelated to some degree. Again, there is a natural physical explanation: if $S_1(t)$ increases, because the magnetosphere is compressed relative to its equilibrium state, then the spin-up component of the accretion torque \citep{Ghosh_1979} exceeds the hydromagnetic spin-down component, and $P_1(t)$ decreases. However, Equation (\ref{eq:SLangevin}) indicates that $S_1(t)$ can increase stochastically as well (e.g.\ due to local hydromagnetic instabilities) and does so independently of the location of the disk-magnetosphere boundary. In the latter circumstances, $S_1(t)$ and $P_1(t)$ are not always anticorrelated; indeed, they may even be positively correlated in some objects, if the hydromagnetic spin-down component of the torque rises sufficiently. 

In the bottom panel of Figure \ref{fig:CorrDistribution}, we find no strong evidence of instantaneous (anti)correlations between $S_1(t)$ and $L_1(t)$, with $-0.1 \lesssim r[S_1(t),L_1(t)] \lesssim 0.1$ in general, except for three cases:  $r[S_1(t),L_1(t)] = 0.13\pm 0.04$ for SXP 11.5, $r[S_1(t), L_1(t)] = 0.20\pm 0.03$ for SXP 59.0,  and $r[S_1(t),L_1(t)] = 0.17 \pm 0.03$ for SXP 18.3.  The results in the bottom panel of Figure \ref{fig:CorrDistribution} are in good agreement with the results for SXP 18.3 in Section 4.2 in \cite{OLeary_2023}. 

\section{Conclusion}\label{Sec:Conc}

Current estimates of $\mu$ for accretion-powered pulsars in the SMC are inferred using time-averaged observables $\bar{P} = \langle P(t) \rangle$ and $\bar{L} = \langle L(t) \rangle$, and rely on a priori assumptions about (i) the rotational state of the star, e.g.\ magnetocentrifugal equilibrium; and (ii) the radiative processes near the disk-magnetosphere boundary, e.g.\ $\bar{\eta} = 1$ \citep{Klus_2014,Ho_2014,Mukherjee_2015,Shi_2015,Dangelo_2017,Karaferias_2023}. The radiative efficiency cannot be inferred uniquely using time-averaged statistics because it is degenerate with the accretion rate.  Hence one must be careful about interpreting $\mu$, when $\bar{\eta}$ is prescribed arbitrarily. Moreover, the RXTE PCA measurements $P(t_n)$ and $L(t_n)$ contain important time-dependent information about the complex accretion physics and radiative processes at the disk-magnetosphere boundary for X-ray pulsars in the SMC and elsewhere, which is lost after averaging over time.

In this paper, we apply the signal processing framework developed by \cite{Melatos_2022} to RXTE PCA $P(t_n)$ and $L(t_n)$ time series for 52 SMC HMXBs, without assuming magnetocentrifugal equilibrium \citep{Melatos_2022,OLeary_2023}. The output of the analysis is subject to conservative acceptance criteria, e.g.\ the one-dimensional $\beta_1$ and $\beta_2$ posterior distributions are unimodal and the inferred radiative efficiency satisfies $0 < \bar{\eta} < 1$. The analysis yields new results for 24 out of 52 accretion-powered pulsars in the SMC.  We estimate the temporal evolution of the hidden variables $Q(t_n)$ and $S(t_n)$ and static parameters $\bm{\Theta} = (\beta_1,\beta_2, \gamma_Q, \gamma_S, \sigma_{QQ}, \sigma_{SS})$ --- and hence $\mu$ and $\bar{\eta}$ --- by applying a nonlinear, unscented Kalman filter \citep{Wan_2000,Wan_2001} to a model that combines the canonical magnetocentrifugal torque with mean-reverting fluctuations in $Q(t)$ and $S(t)$. The nonlinear Kalman filter framework is applied to 10 systems classified as existing in a state of rotational disequilibrium and 14 systems classified as near rotational equilibrium \citep{Yang_2017}. The new results in Sections \ref{Sec:MagPE}--\ref{Sec:Corr}, summarized and itemized below, shed light on four important facets of accretion physics for HMXBs in the SMC.

(i) The analysis points to relaxation rates and rms noise amplitudes satisfying $-8 \lesssim \log_{10}[\gamma_A/(1\, \rm{s^{-1}})] \lesssim -6$ and $-6 \lesssim \log_{10}[\sigma_{AA}\bar{A}^{-1}/(1 \, \rm{s^{-1/2}})] \lesssim -2$ respectively, for $A\in \{Q,S\}$, broadly consistent with observational studies of $P(t)$ and $L(t)$ fluctuations \citep{Bildsten_1997, Mukherjee_2018, Serim_2022}, and in good agreement with the results presented for SXP 18.3 in Section 4 in \cite{OLeary_2023}. It reveals that the characteristic time scales of mean-reversion are shorter for $S$ fluctuations than for $Q$ fluctuations, as seen in the top panel of Figure \ref{fig:ParamDistribution}.

(ii) The analysis yields 24 new estimates of the magnetic dipole moments $\mu$ of HMXBs in the SMC, ranging from a minimum of $ 2.5^{+0.20}_{-0.20} \, \times 10^{30} \, \rm{ G \, cm^3}$ for SXP 8.80 to a maximum of $ 1.1^{+0.11}_{-0.15} \, \times 10^{32} \, \rm{ G \, cm^3}$ for SXP 264. The estimated values are smaller in 21 out of 24 objects than those inferred in \cite{Klus_2014}. The difference between $\mu$ estimated in \cite{Klus_2014} and the present analysis is $\leq 1.2 \, \rm{dex}$; see the last column in Table \ref{tab:KlusMuOverlap} for details.

(iii) The analysis lifts the degeneracy between the accretion rate and radiative efficiency and yields independent estimates of $\bar{\eta}$ for the sample of 24 SMC HMXBs. The inferred radiative efficiency ranges from a minimum of $0.1\%$ to a maximum of $76\%$.  The sample mean equals $12\%$, in line with theoretical estimates, i.e.\ $0.1 \lesssim \bar{\eta} \lesssim 0.2$ \citep{Frank_2002,Longair_2010}. There is new evidence of moderately significant anticorrelations described by a power law of the form $\bar{\eta} \propto \bar{Q}^{-9/10}$ with a coefficient of determination $R_{\bar{\eta}\bar{Q}} = 0.98$, visible in the second panel in Figure \ref{fig:EtaDistribution}. 

(iv) The analysis provides new information about how the scalar Maxwell stress at the disk-magnetosphere boundary $S(t)$, which is notoriously difficult to measure, fluctuates with time. The analysis reveals positive correlations between the hidden variables $Q(t)$ and $S(t)$ for 13 out of the 24 stars analyzed in this paper, i.e.\ we observe spikes in $S(t)$ when the magnetosphere is compressed relative to $\bar{Q}$, i.e.\ $Q(t)>\bar{Q}$. The same correlation occurs in the X-ray transient SXP 18.3 in the fourth and fifth panels of Figure 1 in \cite{OLeary_2023}. We also find evidence of moderate anticorrelations between $S(t)$ and $P(t)$ for 17 SMC HMXBs, e.g.\ in the first and last panels in Figure \ref{fig:AppWEKFTracking} for the X-ray transient SXP 4.78, where we observe a sharp decrease in $S(t)$ for $P(t)>\bar{P}$ between MJD 51600 and MJD 51700, and a spike in $S(t)$ between MJD 54400 and MJD 54600, accompanied by a sharp decrease in $P(t)$.  

The next steps are as follows. (i) Calibrate the Kalman filter output using samples of accreting millisecond pulsars, e.g.\ SAX J1808.4$-$3658, IGR J00291$+$5934, and XTE J1751$-$305 \citep{Patruno_2021}, as well as persistent galactic X-ray pulsars, e.g.\ Her X$-$1, 4U 1626$-$67, and OAO 1657$-$415, whose magnetic field strengths, and hence magnetic dipole moments, are measured via cyclotron resonant scattering features \citep{Staubert_2019} or otherwise \citep{Hartman_2008}. For example, the Fermi Gamma-Ray Burst Monitor Accreting Pulsar Program\footnote{\href{https://gammaray.msfc.nasa.gov/gbm/science/pulsars.html}{https://gammaray.msfc.nasa.gov/gbm/science/pulsars.html}} publishes $P(t_n)$ time series -- but not aperiodic X-ray flux -- for eight persistent galactic X-ray pulsars, with   $700 \lesssim N \lesssim 2600$ samples. One needs a formal nonlinear identifiability analysis \citep{bellman_1970} to assess the subset of parameters that can be estimated, given $P(t_n)$ measurements only. (ii) Search the time-dependent fastness parameters $[R_{\rm m}(t)/R_{\rm c}(t)]^{3/2}$ of the 24 SMC HXMBs analyzed here for signatures of Rayleigh-Taylor unstable accretion regimes as observed in global, three-dimensional magnetohydrodynamic numerical simulations \citep{Romanova_2014,Romanova_2015,Blinova_2016}. We will explore these opportunities in forthcoming papers.

The authors thank Filippo Anzuini, Julian Carlin, Nathalie Deganaar, Konstantinos Kovlakas, Alessandro Patruno, Phil Uttley, Andr\'es Vargas, and Rudy Wijnands for helpful discussions about Bayesian inference, X-ray timing data, and high energy astrophysics. We also acknowledge a helpful discussion with Robin Evans regarding unscented Kalman filters. This research was supported by the Australian Research Council Centre of Excellence for Gravitational Wave Discovery, grant number CE170100004. NJO’N is the recipient of a Melbourne Research Scholarship. DMC acknowledges funding through the National Science Foundation Astronomy and astrophysics research grant 2109004.  DMC,
SB, and STGL acknowledge funding through the National Aeronautics and Space Administration Astrophysics Data Analysis Program grant NNX14-AF77G.

\bibliographystyle{aasjournal}
\bibliography{main.bib}

\appendix

\section{Parameter estimation with an unscented Kalman filter} \label{AppA:UKF}

The unscented Kalman filter (UKF) \citep{Julier_1997} is a signal processing tool employed in nonlinear state and parameter estimation problems; see Section 1 in \cite{Wan_2000} and
Chapter 2 in \cite{Challa_2011} for a comparison of nonlinear Kalman filters, and Chapters 7 and 19 in \cite{Zarchan_2005} for a practical guide on their implementation. The UKF is a recursive Bayesian filter. The goal of nonlinear recursive Bayesian estimation with a UKF is to compute the posterior density of the hidden states $\bm{X}(t_n)$, given a set of noisy or incomplete observations $\bm{Y}$. The algorithm proceeds sequentially, i.e.\ the epoch of the estimated state is updated with each new observation, and all information obtained from previous measurements is contained in the current state and covariance estimates, denoted by $\bm{X}$ and $\bm{P}$, respectively. 

In this Appendix, we summarize for the convenience of the reader the UKF algorithm for parameter estimation problems using the nested sampling algorithm; see Sections 4 and 5 in \cite{Ashton_2022} and \cite{Skilling_2006} for further details. In Appendix \ref{AppA:TheUT}, we introduce the unscented transform as a technique to estimate the first and second order moments of a  random variable, subject to a nonlinear transformation. The UKF algorithm for state tracking is summarized in Appendix \ref{AppA:StateTracking}. A brief overview of static parameter estimation using the Kalman filter log-likelihood function and a nested sampling algorithm is given in Appendix \ref{AppA:PE}. The astrophysical priors as well as the nested sampler settings used in the analysis above are summarized briefly in Appendices \ref{App:Priors} and \ref{App:SamplerSettings}, respectively. The algorithm in this appendix together with the astrophysical priors and nested sampler settings can be used to reproduce independently the results in the body of the paper, e.g.\ Figures \ref{fig:ParamDistribution}--\ref{fig:CorrDistribution}.
\subsection{Unscented transform}\label{AppA:TheUT}
The UKF shares the same predictor-corrector design as the linear Kalman filter; see Algorithms 1 and 3 in \cite{Challa_2011}.
The key difference between the UKF and other nonlinear Bayesian filters, e.g.\ the extended Kalman filter, stems from the manner in which the state distribution, approximated as a Gaussian random variable, is represented for propagation through the nonlinear system dynamics \citep{Wan_2001}. 

In the context of nonlinear Bayesian estimation, the unscented transform adopts a deterministic sampling approach using a set of weighted sample  points (``sigma vectors'') to calculate the first- and second-order moments of the $p$-dimensional random vectors, $\bm{X}$ and  $\bm{Z} = F(\bm{X})$. The algorithm proceeds as follows. Let $\bm{\mathcal{X}}$ denote an $s\times p$ matrix of $s = 2p + 1$ nonuniformly spaced, deterministic sigma vectors $\bm{\mathcal{X}}_1,\hdots,\bm{\mathcal{X}}_s$, and $w^{\rm{m}}_1,\hdots,w^{\rm{m}}_s$ and $w^{\rm{c}}_1,\hdots,w^{\rm{c}}_s$ denote the corresponding scalar weights. The superscripts ``m'' and ``c'' denote scalar weights associated with the mean and covariance of a random vector, respectively. The sigma vectors $\bm{\mathcal{X}}_i$ and weights $w^{\rm{m}}_i$ and $w^{\rm{c}}_{i}$ for $i=1,\hdots,s$, are defined to satisfy 
\begin{equation}
    \langle \bm{X} \rangle = \sum^{s}_{i=1} w^{\rm{m}}_i \bm{\mathcal{X}}_i,
\end{equation}
and 
\begin{equation}
    \langle \bm{X} \, \bm{X}^{\tn{T}} \rangle = \sum^{s}_{i=1} w^{\rm{c}}_i \, [\bm{\mathcal{X}}_i - \langle \bm{X} \rangle][\bm{\mathcal{X}}_i - \langle \bm{X} \rangle]^{\tn{T}},
\end{equation}
i.e.\ the first two sample moments of the sigma points match the true mean and covariance of the prior random vector $\bm{X}$ \citep{Challa_2011}. Tables I and II in \cite{Menegaz_2015} provide a summary of $\bm{\mathcal{X}}_i, w^{\rm{m}}_i,$ and $w^{\rm{c}}_i$ prescriptions. In this paper, we adopt the standard symmetric set of weighted sigma points in \cite{Wan_2001}; see Table I in \cite{Menegaz_2015} for more details.

The unscented transform approximates the first and second order moments of $\bm{Z}$ using a weighted sum of nonlinear function evaluations $\bm{\mathcal{Z}} = \bm{F}(\bm{\mathcal{X}})$, where $\bm{\mathcal{Z}}$ denotes an $s \times p$ matrix of propagated sigma points $\bm{\mathcal{Z}}_1,\hdots,\bm{\mathcal{Z}}_s$, with
\begin{equation} \label{Eq:Z2Mean}
    \langle \bm{Z} \rangle \approx \sum^{s}_{i=1} w^{\rm{m}}_i \bm{\mathcal{Z}}_i,
\end{equation}
and 
\begin{equation} \label{Eq:Z2Cov}
    \langle \bm{Z} \, \bm{Z}^{\tn{T}} \rangle \approx \sum^{s}_{i=1} w^{\rm{c}}_i [\bm{\mathcal{Z}}_i - \langle \bm{Z} \rangle][\bm{\mathcal{Z}}_i - \langle \bm{Z} \rangle]^{\tn{T}}.
\end{equation}
Similarly, cross-correlations $\langle \bm{X} \, \bm{Z}^{\tn{T}} \rangle$ are approximated analogously to Equation (\ref{Eq:Z2Cov}), viz. 
\begin{equation} \label{Eq:Z2CrossCorr}
    \langle \bm{X} \bm{Z}^{\tn{T}} \rangle \approx \sum^{s}_{i=1} w^{\rm{c}}_{i}[\bm{\mathcal{X}}_i - \langle \bm{X} \rangle][\bm{\mathcal{Z}}_i - \langle \bm{Z}\rangle]^{\tn{T}}.
\end{equation}
Equations (\ref{Eq:Z2Mean})--(\ref{Eq:Z2CrossCorr}) approximate the posterior mean and covariance of $\bm{Z}$ to second order \citep{Challa_2011}; see Appendix A in \cite{Wan_2001} for a summary of the approximation's accuracy.

\subsection{State tracking}\label{AppA:StateTracking}

The UKF is a direct extension of the unscented transform for recursive estimation. In this paper, we consider a system of $p$ first-order, noise-driven, nonlinear differential equations of the form
\begin{equation}\label{Eq:DENL}
    \frac{\tn{d} \bm{X}}{\tn{d}t} = \bm{F}(t, \bm{X}, \bm{\Theta}) + \bm{\xi}(t),
\end{equation}
where $\bm{F}(t, \bm{X}, \bm{\Theta})$ denotes a time-dependent nonlinear function of the hidden states $\bm{X}(t)$ and a set of static parameters $\bm{\Theta}$. In the accretion-powered pulsar context, the  components of $\bm{F}(t, \bm{X}, \bm{\Theta})$ are defined by the right-hand sides of Equations (\ref{Eq:SpinEquation})--(\ref{eq:SLangevin}), respectively.   The  vector components of $\bm{\xi}(t)$ are white noise, zero-mean driving terms, which satisfy the ensemble statistics
\begin{eqnarray}
    \langle \bm{\xi}_j(t) \rangle &=& 0, \\ \langle \bm{\xi}_j(t) \bm{\xi}_k(t')^{\rm{T}} \rangle &=& \delta_{jk} \bm{\sigma}^2 \delta(t - t').
\end{eqnarray}
The superscript `T' denotes the matrix transpose, and the Kronecker and delta functions are denoted by $\delta_{jk}$ and $\delta(t - t')$ respectively, for $j,k = 1,\hdots,p$. 

In the context of magnetocentrifugal accretion, the nonlinear differential equations implicitly defined by the noiseless terms on the right-hand side of Equation (\ref{Eq:DENL}) do not exhibit known analytical solutions in the disequilibrium case, whereas the noiseless, linear system of differential equations associated with the equilibrium scenario is solved using standard techniques; see Equation (A3) in \cite{OLeary_2023}. Accordingly, Equation (\ref{Eq:DENL}) gives rise to the recurrence relation
\begin{equation}\label{Eq:UKFRecurrence}
\bm{X}_n = \bm{G}_{n-1} + \bm{\eta}_{n},
\end{equation}
where $\bm{G}_{n-1} = \bm{G}(t_{n-1},\bm{X}_{n-1},\bm{\Theta})$ is given by
\begin{equation}\label{Eq:DefinitionofG}
\bm{G}(t_{n-1},\bm{X}_{n-1},\bm{\Theta}) = \bm{X}_{n-1} + \int^{t_n}_{t_{n-1}}  \tn{d}t' \bm{F}(t', \bm{X}, \bm{\Theta}), 
\end{equation}
and the additive noise terms 
\begin{equation}
    \bm{\eta}_{n} \approx \int^{t_{n}}_{t_{n-1}}  \tn{d}t' \exp{[\nabla_n \bm{F} (t_n - t')]} \bm{\xi}(t'),
\end{equation}
are normally distributed, zero-mean vectors with covariance $\bm{Q}$, i.e.\ $\bm{\eta}\sim \mathcal{N}(\bm{0},\bm{Q})$ and $\nabla_n \bm{F}$ denotes the Jacobian matrix of $\bm{F}$ evaluated at $\bm{X}^{-}_{n}$; see Equation (\ref{Eq:ApproximateMean}) below for details on how $\bm{X}^{-}_{n}$ is approximated using the unscented transform. The latter prescription for $\bm{\eta}_{n}$ is standard in nonlinear state and parameter estimation problems using an unscented (or extended) Kalman filter \citep{Svensson_2019}; see also Chapter 12 in \cite{Gustafsson_2010} for a step-by-step guide on implementing the aforementioned approach in practice, as well as overviews of other, similar but different techniques to approximate $\bm{\eta}_{n}$ in nonlinear filtering problems.

The state variables $\bm{X}(t)$ are not observed directly. They are related to a set of noisy observations $\bm{Y}$, viz.
\begin{equation}\label{Eq:NLMeasurement}
\bm{Y}_n = \bm{C}[\bm{X}(t_n)] + \bm{N}(t_n),
\end{equation}
where $\bm{C}(\bm{X})$ denotes a nonlinear function of the state variables $\bm{X}(t_n)$, defined in terms of the right-hand sides of Equations (\ref{Eq:SpinPeriodMain}) and (\ref{Eq:LuminosityMain}), respectively.  The additive noise terms in Equation (\ref{Eq:NLMeasurement}) are zero-mean, Gaussian random variables with covariance $\bm{\Sigma}$, i.e.\ $\bm{N}\sim \mathcal{N}(0,\bm{\Sigma})$. 

Given the system of nonlinear stochastic differential equations (\ref{Eq:DENL}), its associated solution (\ref{Eq:UKFRecurrence}), and  a set of noisy or incomplete observations $\bm{Y}_n$, a UKF can be applied to perform state estimation. In its simplest form, state tracking with a UKF involves predict and update stages.  At time $t_n$, the predicted mean and covariance of $\bm{X}_n$ are given explicitly by the first- and second-order moments of Equation (\ref{Eq:UKFRecurrence}), viz.
\begin{equation}\label{Eq:ExplicitExpectationMean}
    \langle \bm{X}_n \, | \, \bm{Y}_{n-1} \rangle = \langle \bm{G}_{n-1} \, | \, \bm{Y}_{n-1} \rangle,
\end{equation}
and 
\begin{equation} \label{Eq:ExplicitExpectationCov}
        \langle \bm{X}_n^{} \, \bm{X}_n^{\tn{T}} \, | \, \bm{Y}_{n-1} \rangle = \langle [\bm{G}_{n-1} - \langle \bm{G}_{n-1} \rangle] \, [\bm{G}_{n-1} - \langle \bm{G}_{n-1} \rangle]^{\tn{T}} \, | \, \bm{Y}_{n-1} \rangle + \langle \bm{\eta}_n^{} \bm{\eta}_n^{\tn{T}} \rangle,
\end{equation}
where $\langle \bm{X}_n | \bm{Y}_{n-1} \rangle$ denotes the expected value of $\bm{X}$ at time $t_n$ given the measurements $\bm{Y}$ up to time $t_{n-1}$. The UKF approximates the right-hand sides of Equations (\ref{Eq:ExplicitExpectationMean}) and (\ref{Eq:ExplicitExpectationCov}) using the unscented transform in the predict stage as follows. We introduce an $s \times p$ matrix of numerically propagated sigma vectors $\bm{\mathcal{X}}_{n} = \bm{G}(t_{n-1},\bm{\mathcal{X}}_{n-1},\bm{\Theta})$, where $\bm{\mathcal{X}}_{n-1}$ are calculated to match $\bm{X}_{n-1}$ and $\bm{P}_{n-1}$ at time $t_{n-1}$. The predicted mean $\bm{X}^{-}_n$ and covariance $\bm{P}^{-}_{n}$ then follow respectively from Equations (\ref{Eq:Z2Mean}) and (\ref{Eq:ExplicitExpectationMean}), and Equations (\ref{Eq:Z2Cov}) and (\ref{Eq:ExplicitExpectationCov}), viz.
\begin{equation}\label{Eq:ApproximateMean}
    \bm{X}^{-}_n \approx \sum^s_{i=1} w^{\rm{m}}_i \bm{\mathcal{X}}_{i,n},
\end{equation}
and 
\begin{equation}\label{Eq:ApproximateCov}
 \bm{P}^{-}_{n} \approx \bm{Q}_n + \sum^{s}_{i=1} w^{\rm{c}}_i (\bm{\mathcal{X}}_{i,n} - \bm{X}^{-}_n)(\bm{\mathcal{X}}_{i,n} - \bm{X}^{-}_n)^{\tn{T}}.
\end{equation} 

The update stage then uses the measurement at time $t_n$, with a new set of sigma vectors $\bm{\mathcal{X}}_{i,n}$,\footnote{The sigma vectors $\bm{\mathcal{X}}_{i,n}$ are recalculated in the update stage to match the estimated mean $\bm{X}^{-}_n$ and covariance $\bm{P}^{-}_n$ in Equations (\ref{Eq:ApproximateMean}) and (\ref{Eq:ApproximateCov}), respectively.} to update the state and covariance, viz. 
\begin{equation}\label{Eq:MeanUpdate}
    \bm{X}_n = \bm{X}^{-}_{n} + \bm{k}_n (\bm{Y}_n - \bm{Y}^{-}_{n}),
\end{equation}
and 
\begin{equation} \label{Eq:CovUpdate}
    \bm{P}_n = \bm{P}^{-}_n - \bm{k}_n \bm{\Psi}^{\tn{T}}_{n}.
\end{equation}
In Equation (\ref{Eq:MeanUpdate}), the predicted measurement $\bm{Y}^{-}_n$ is calculated analogously to Equation (\ref{Eq:ApproximateMean}), i.e.\ we construct a matrix $\bm{\mathcal{Y}} = \bm{C}(\bm{\mathcal{X}})$ of $s = 2p +1$ sigma vector measurement estimates $\bm{\mathcal{Y}}_1,\hdots,\bm{\mathcal{Y}}_s$, and approximate $\bm{Y}^{-}_{n}$ using a weighted sum according to 
\begin{equation} 
    \bm{Y}^{-}_n = \sum_{i=1}^{s} w^{\rm{m}}_i \bm{\mathcal{Y}}_{i,n}.
\end{equation}
The Kalman gain,
\begin{equation} \label{Eq:KalmanGain}
\bm{k}_n = \bm{\Psi}_{n} \bm{s}^{-1}_n,
\end{equation}
in Equations (\ref{Eq:MeanUpdate}) and (\ref{Eq:CovUpdate}) is defined in terms of the predicted state and measurement cross covariance $\bm{\Psi}_n$, and the innovation covariance $\bm{s}_n$ viz.
\begin{equation}
    \bm{\Psi}_n = \sum_{i=1}^{s} w^{\rm{c}}_i(\bm{\mathcal{X}}_{i,n} - \bm{X}^{-}_n)(\bm{\mathcal{Y}}_{i,n} - \bm{Y}^{-}_n)^{\tn{T}},
\end{equation}
and 
\begin{equation} \label{Eq:InnovationCovariance}
    \bm{s}_n = \bm{\Sigma}_n +  \sum^{s}_{i=1} w^{\rm{c}}_i (\bm{\mathcal{Y}}_{i,n} - \bm{Y}^-_{n})(\bm{\mathcal{Y}}_{i,n} - \bm{Y}^-_{n})^{\tn{T}}.
\end{equation}

The hidden state recursion equation, Equation (\ref{Eq:MeanUpdate}), has three key features. (i) It is linear and shares the same mathematical structure across linear, extended, and unscented Kalman filters, as well as extended and unscented particle filters \citep{Wan_2001}. (ii) The recursive update of $\bm{X}_n$ involves a mixture of dynamical and measurement information, and the relative weights accorded to the two types of information depend on the respective uncertainties, quantified by $\bm{Q}_n$ and $\bm{\Sigma}_n$. (iii) It yields the minimum mean-squared error estimate of $\bm{X}_n$ assuming Gaussian statistics.

\subsection{Parameter estimation}\label{AppA:PE}

In this paper, we combine the Kalman filter log-likelihood \citep{Meyers_2021}
\begin{equation}
   \log p (\{\bm{Y}_n\}| \bm{\Theta}) = \sum^{N}_{n=1} \mathcal{N}(\bm{Y}_n^{-},\bm{s}_n|\bm{\Theta}),
\end{equation}
with the prior distribution $p(\bm{\Theta})$ to estimate the posterior on the parameters $\bm{\Theta}$ according to Bayes' rule,
\begin{equation} \label{Eq:BayesRule}
    p(\bm{\Theta}| \{\bm{Y}_n\}) = \frac{p (\{\bm{Y}_n\}| \bm{\Theta}) \, p(\bm{\Theta})}{Z(\{\bm{Y}_n\})},
\end{equation}
where the denominator on the right-hand side of Equation (\ref{Eq:BayesRule}) is known as the marginalized likelihood and is defined according to
\begin{equation} \label{eq:evidence}
    Z \lr{  \{\bm{Y}_n\} } = \int \textnormal{d}\bm{\Theta} \, p(\{\bm{Y}_n\}|\bm{\Theta}) p\lr{\bm{\Theta}} .
\end{equation}

The \texttt{dynesty} nested sampler \citep{Speagle_2020} is employed to numerically approximate Equation (\ref{eq:evidence}). The steps of nested sampling are laid out in Section \ref{Sec:UKF}.

\subsection{Astrophysical priors} \label{App:Priors}

The nested sampler ingests as input the prior distribution $p(\bm{\Theta})$ for the six static parameters $\bm{\Theta} = (\beta_1, \beta_2, \gamma_Q, \gamma_S, \sigma_{QQ}, \sigma_{SS})$. We assume a log-uniform prior  distribution $\log_{10} \mathcal{U}(a,b)$ for each parameter bound between $a$ (lower) and $b$ (upper). The priors used in this paper are summarized in Table \ref{tab:EqParams}. Their astrophysical motivation is as follows. For systems accreting via a disk, observational studies of $L(t)$ and $P(t)$ fluctuations point to relaxation processes operating on time scales of days to weeks \citep{bildsten_1998,Mukherjee_2018,Serim_2022,Melatos_2022}. Accordingly we adopt $-8 \leq \log_{10}[\gamma_A/(1 \, \rm{s}^{-1})] \leq -5$ for $A \in \{Q, S \}$ as a typical range for the mean-reversion time scales. We adopt broad priors for the rms noise amplitudes, viz. $-6 \leq \log_{10}[\sigma_{AA} \bar{A}^{-1}/(1 \, \rm{s}^{-1/2})] \leq -1$ for $A \in \{Q, S \}$. Given that this is the first time $\beta_1$ and $\beta_2$ are measured with a Kalman filter, and in the absence of observational constraints on their expected range, we set broad, uninformative priors on $\beta_1$ and $\beta_2$, viz.\   $-12 \leq \log_{10}[\beta_1/(1 \, \rm{s}^{-1})] \leq -7$ and $-12 \leq \log_{10}[\beta_2/(1 \, \rm{s}^{-1})] \leq -7$. In future studies it may be appropriate to adopt more informative priors as more accretion-powered pulsars are analyzed with next-generation X-ray telescopes, yielding larger sample sizes $N$ per source as well as revealing a more refined picture of the distribution of (for example) $\beta_1$ and $\beta_2$.

\begin{table}
\centering
 \begin{tabular}{c c c} 
 \hline
Parameter & Units  & Prior \\ [1ex] \hline 
$\beta_1$ & $\rm{s}^{-1}$ & $\log \mathcal{U} (10^{-12}, 10^{-7})$ \\ [1ex] 
$\beta_2$ & $\rm{s}^{-1}$ & $\log \mathcal{U} (10^{-12}, 10^{-7})$  \\  [1ex]
$\gamma_Q$ & $\rm{s}^{-1}$ & $\log \mathcal{U} (10^{-8}, 10^{-5})$   \\  [1ex]
$\gamma_S$ & $\rm{s}^{-1}$ & $\log \mathcal{U} (10^{-8}, 10^{-5})$   \\  [1ex]
$\sigma_{QQ}/\bar{Q}$ & $\rm{s}^{-1/2}$ & $\log \mathcal{U} (10^{-6}, 10^{-1})$   \\  [1ex]
$\sigma_{SS}/\bar{S}$ & $\rm{s}^{-1/2}$ & $\log \mathcal{U} (10^{-6}, 10^{-1})$   \\  [1ex]
 \hline
 \end{tabular}
 \caption{Prior distribution $p(\bm{\Theta})$ ingested by the \texttt{dynesty} nested sampler. We assume a log-uniform distribution for each parameter, denoted $\log \mathcal{U}(a,b)$, where $a$ and $b$ denote the lower and upper prior bounds, respectively.}
 \label{tab:EqParams}
\end{table}

\subsection{Nested sampler settings}\label{App:SamplerSettings}

The results in Section \ref{Sec:MagPE}--\ref{Sec:Corr} are insensitive to the \texttt{dynesty} nested sampler settings, whose main tunable features are the number of live points $N_{\rm{live}}$, discussed in Section \ref{Sec:UKF} above, and a suitable stopping criterion $\Delta$. We repeated the analysis outlined in Section \ref{Sec:UKF} for a subsample of the X-ray pulsars in Figure \ref{fig:ObsDistribution} using $250$ live points, $500$ live points, and $1000$ live points, and obtained comparable posterior distributions for the six model parameters specified in Table \ref{tab:EqParams} (results not listed for brevity). In this paper, we adopt $N_{\rm{live}} = 500$ and $\Delta = 0.1$. 

\section{An example of Kalman filter data products: SXP $4.78$}\label{App:WorkedExample}

The Kalman filter yields a number of data products for the 24 SMC HMXBs analyzed in this paper. Here we summarise the analysis products for a single, exemplary system, namely SXP 4.78, classified in Table 3 in \cite{Yang_2017} as being in magnetocentrifugal disequilibrium, with $\tn{d}P/\tn{d}t < 0$. The results in this appendix are included to assist the interested reader with reproducibility. 

\begin{figure}
\centering{
\hspace*{-2.0cm}
\vspace*{-0.5cm}
    \includegraphics[width=0.80\textwidth, keepaspectratio]{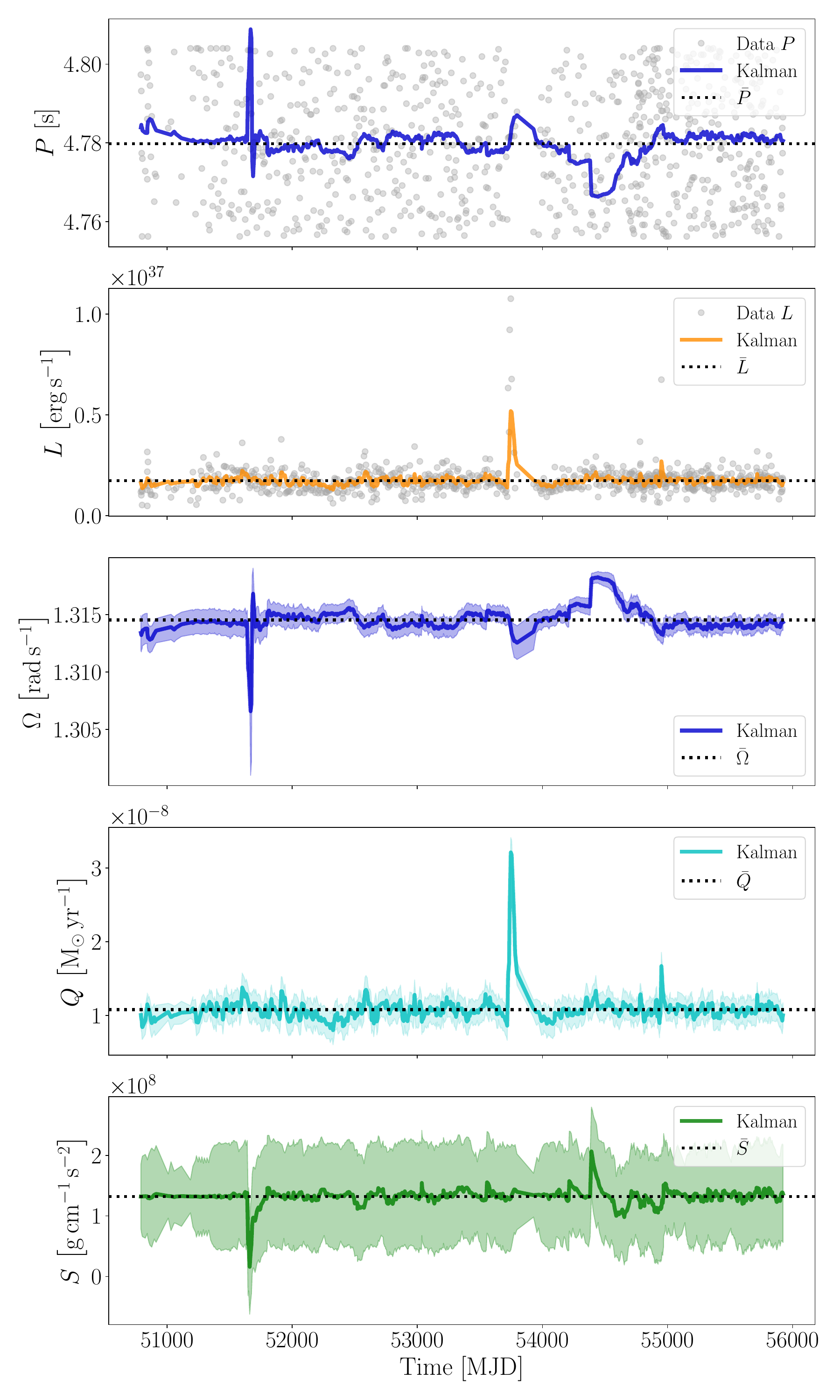}}
        \caption{Kalman filter state tracking applied to the exemplary object SXP 4.78. Inputs: RXTE PCA measurements of pulse period fluctuations $P_n$ (units: $\rm{s}$; first panel, gray points) and aperiodic X-ray luminosity fluctuations $L_n$ (units: $\rm{erg \, s^{-1}}$; second panel, gray points). Outputs: Kalman filter estimates $P^{-}_n$ (top panel, blue curve) and $L^{-}_n$ (second panel, orange curve), and the state variables $\Omega_n$ (units: $\rm{rad\, s^{-1}}$; third panel, blue curve), $Q_n$ (units: $\rm{M_{\odot} \, yr^{-1}}$; fourth panel, cyan curve) and $S_n$ (units: $\rm{g \, cm^{-1} \, s^{-2}}$; fifth panel, green curve). In each panel, $\bar{P}$, $\bar{L}$, $\bar{\Omega}$, $\bar{Q}$, and $\bar{S}$ are overplotted as black, dashed lines. The quantities $\bar{P},\bar{L},\bar{\Omega}$ are calculated directly from the data sample means. The parameters $\bar{Q}$ and $\bar{S}$ are inferred from $\beta_1$ and $\beta_2$, discussed in Section \ref{SubSec:MagMomentObs}. The shaded regions correspond to the 1-$\sigma$ state estimate uncertainties, returned by the UKF. The time units on the horizontal axis are MJD. }
    \label{fig:AppWEKFTracking}
\end{figure}

In Figure \ref{fig:AppWEKFTracking} we summarise the Kalman filter inputs and outputs as functions of time. In the top two panels, the RXTE PCA $P(t_n)$ and $L(t_n)$ measurements are plotted as gray points, and the Kalman filter estimates are overplotted as blue and orange solid curves, respectively. The black, dashed  lines in the top two panels correspond to $\bar{P}$ and $\bar{L}$, calculated from the sample means of the time series $P(t_n)$ and $L(t_n)$, respectively. The Kalman filter state estimates $\bm{X}_n = [\Omega(t_n), Q(t_n), S(t_n)]$ are plotted as functions of $t_n$ in the bottom three panels of Figure \ref{fig:AppWEKFTracking} as colored, solid curves. The shaded, colored regions correspond to the 1-$\sigma$ state uncertainties, calculated from the square root of the diagonal entries of the state error covariance $\bm{P}_n$. Overplotted as black, dashed  lines are the Kalman filter estimates for $\bar{\Omega}$, $\bar{Q}$, and $\bar{S}$. The latter two parameters are inferred from the magnetocentrifugal parameters $\beta_1$ and $\beta_2$, and $\bar{\Omega}$ is calculated directly from Equation (\ref{Eq:SampleMeansOmega}). The mean-reverting nature of the hidden states $Q(t_n)$ and $S(t_n)$ is evident upon inspection.

\begin{figure}
\centering{
\hspace*{-2cm}
    \includegraphics[width=1.2\textwidth, keepaspectratio]{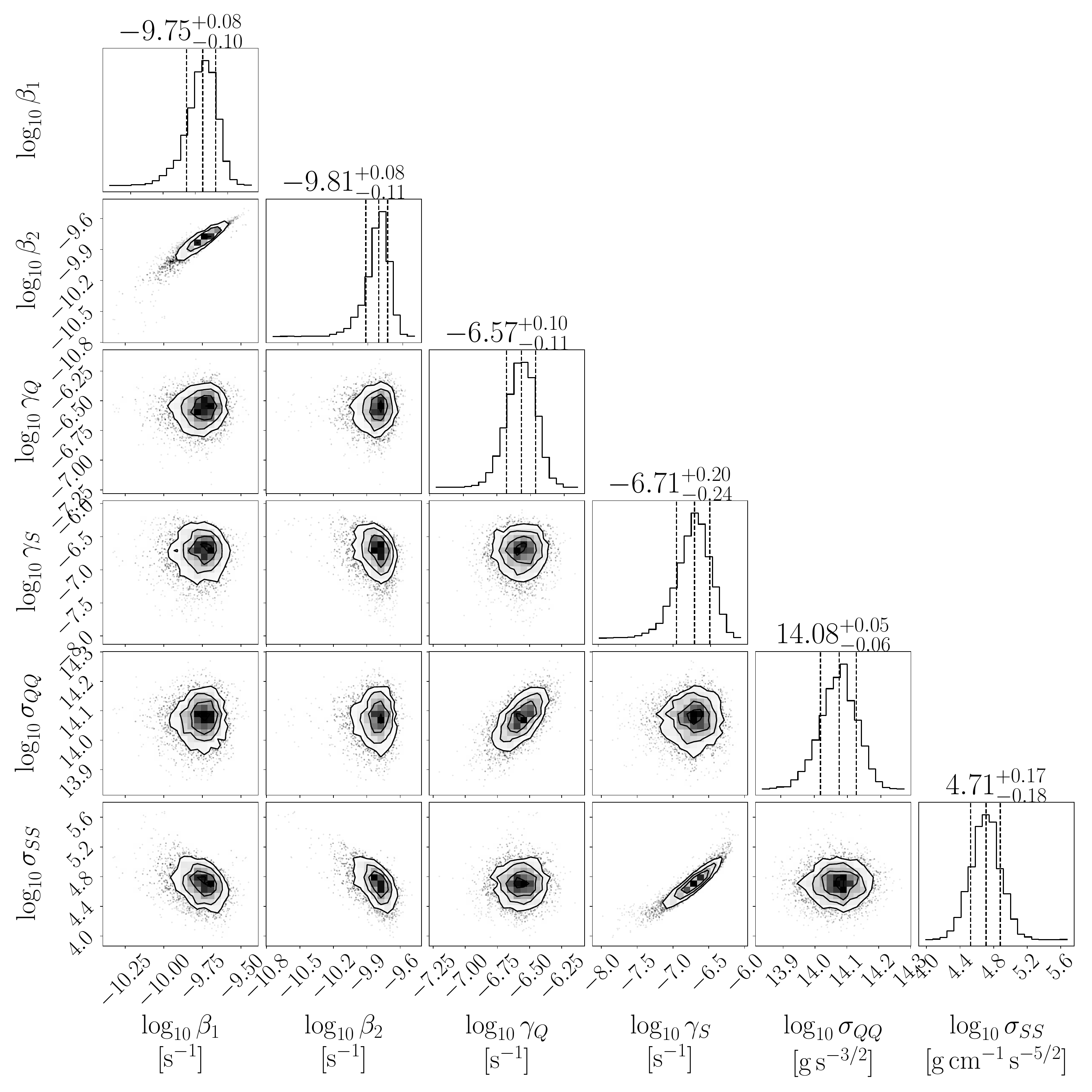}}
    \caption{Corner plot of the posterior distribution of the six fundamental model parameters $\mathbf{\Theta}$ = $(\beta_1, \beta_2, \gamma_A, \sigma_{AA})$, with $A \in \left\{ Q, S \right\}$, for SXP 4.78. The contour plots depict the posterior marginalized over four out of six parameters. The one-dimensional histograms depict the posterior marginalized over five out of six parameters. The distributions are plotted on $\log_{10}$ scales. }
    \label{fig:AppWECornerPlot}
\end{figure}

In Figure \ref{fig:AppWECornerPlot} we present the six-dimensional posterior distribution of the model parameters $\bm{\Theta} = (\beta_1, \beta_2, \gamma_Q, \gamma_S, \sigma_{QQ}, \sigma_{SS})$, returned by the nested sampler from the data. The posterior is visualised in cross-section using a traditional corner plot. The nominal value reported at the top of the one-dimensional posteriors correspond to the posterior median, and the uncertainty in each estimate is quantified using a 68$\%$ credible interval, visible in Figure \ref{fig:AppWECornerPlot} as three, vertical, dashed lines. We employ the posterior maximum likelihood estimate of $\bm{\Theta} = (\beta_1, \beta_2, \gamma_Q, \gamma_S, \sigma_{QQ}, \sigma_{SS})$ returned by the nested sampler to generate the hidden state sequence $\bm{X}_n$, reported in Figure \ref{fig:AppWEKFTracking}. The one-dimensional posteriors are unimodal and all six parameters are estimated unambiguously. There is evidence of correlations in the $\beta_1$-$\beta_2$ and $\gamma_S$-$\sigma_{SS}$ planes for all 24 stars analyzed in this paper. 

\begin{figure}
\centering{
\hspace*{-2.2cm}
    \includegraphics[width=1.2\textwidth, keepaspectratio]{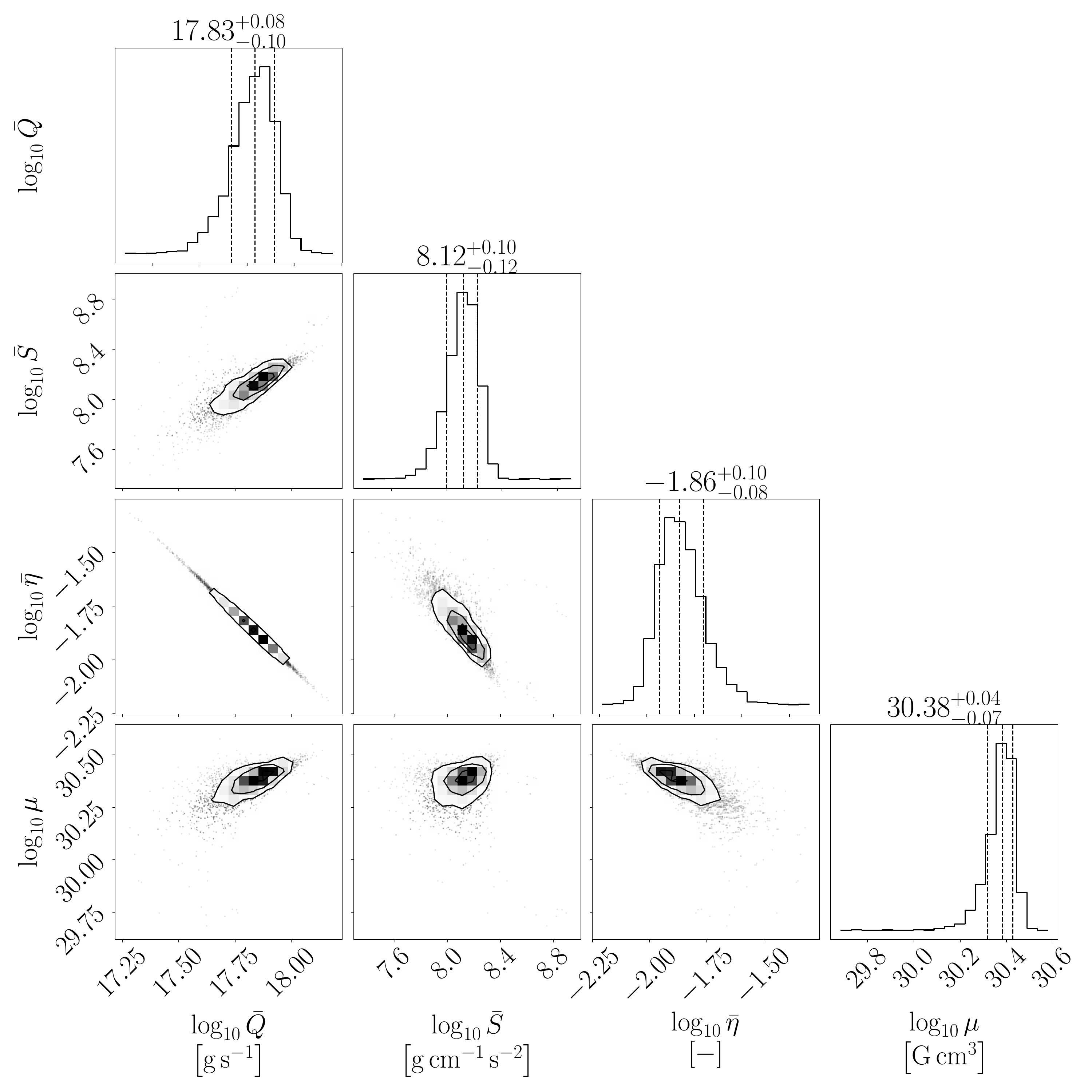}}
    \caption{Corner plot of the posterior distribution of the four inferred static parameters $(\bar{Q},\bar{S},\bar{\eta},\mu)$ for SXP 4.78.  The contour plots depict the posterior marginalized over two out of four parameters. The one-dimensional histograms depict the posterior marginalized over three out of four parameters. The distributions are plotted on $\log_{10}$ scales.}
    \label{fig:AppWECornerPlotParams}
\end{figure}

The four-dimensional posterior distribution of the parameters $(\bar{Q},\bar{S},\bar{\eta},\mu)$ is visualised through a traditional corner plot in Figure \ref{fig:AppWECornerPlotParams}. The histograms are generated by drawing random samples from the marginalized posteriors $\beta_1$ and $\beta_2$ and the normal distribution $A \sim \mathcal{N}(\bar{A},N^{-1} \sigma_{AA}^2 )$, with $A\in \{P, L\}$, where $N^{-1/2}\sigma_{AA}$ denotes the standard error on the sample means $\bar{P}$ and $\bar{L}$. The procedure to generate the histograms in Figure \ref{fig:AppWECornerPlotParams} is described in detail in Section 4.3 in \cite{OLeary_2023}. The one-dimensional posteriors are unimodal and the four parameters $(\bar{Q},\bar{S},\bar{\eta},\mu)$ are estimated unambiguously. There is evidence of anticorrelations in the $\bar{\eta}$-$\bar{Q}$, $\bar{\eta}$-$\bar{S}$, and $\bar{\eta}$-$\mu$ planes for the 24 stars analyzed in this paper.

We refer the reader to Sections 4 in \cite{Melatos_2022} and \cite{OLeary_2023} for detailed discussions on interpreting the astrophysical implications of Figures \ref{fig:AppWEKFTracking}--\ref{fig:AppWECornerPlotParams}.  The purpose of this appendix is simply to record the intermediate data products for an exemplary object to assist reproducibility and provide context for the overall, 24-object analysis.

\end{document}